\documentclass[fleqn,usenatbib]{mnras}

\usepackage{newtxtext,newtxmath}
\usepackage[T1]{fontenc}
\usepackage{ae,aecompl}

\usepackage{graphicx} % Including figure files
\usepackage{amsmath}  % Advanced maths commands
\usepackage{amssymb}  % Extra maths symbols
\usepackage{hyperref}
\usepackage{longtable}
\usepackage{newtxtext,newtxmath}
\usepackage{booktabs}
\usepackage{morefloats}
\usepackage{lipsum}
\usepackage{multirow}
\usepackage{footnote}
\usepackage{multirow}
\usepackage{color}
\definecolor{referee}{rgb}{0.0,0,0}
\usepackage{pdflscape}
\usepackage{ragged2e}
\title[ \text{[C\,{\sc ii}]} 158$\mu$m and warm dust at $z$ = 8.31]{ALMA uncovers the [C\,{\sc ii}] emission and warm dust continuum in a $z=8.31$ Lyman break galaxy}

\author[Tom Bakx]{Tom J. L. C. Bakx$^{1,2}$\thanks{E-mail: bakx@a.phys.nagoya-u.ac.jp (Nagoya University)},
Yoichi Tamura$^{1}$,
Takuya Hashimoto$^{3, 4, 2}$,
Akio K.\ Inoue $^{5,6}$,\newauthor
Minju M. Lee$^{7}$,
Ken Mawatari$^{4,7}$,
Kazuaki Ota$^{9}$,
Hideki Umehata$^{10, 11, 12}$,\newauthor
Erik Zackrisson$^{13}$,
Bunyo Hatsukade$^{12}$, % [0000-0002-2419-3068]
Kotaro Kohno$^{12,14}$, 
Yuichi Matsuda$^{2, 15}$, \newauthor
Hiroshi Matsuo$^{2, 15}$, 
Takashi Okamoto$^{16}$,
Takatoshi Shibuya$^{17}$,
Ikkoh Shimizu$^{2}$, \newauthor 
Yoshiaki Taniguchi$^{9}$,
Naoki Yoshida$^{18,19}$
\\
$^{1}$ Division of Particle and Astrophysical Science, Graduate School of Science, Nagoya University, Aichi 464-8602, Japan.\\
$^{2}$ National Astronomical Observatory of Japan, 2-21-1, Osawa, Mitaka, Tokyo 181-8588, Japan.\\
$^{3}$ Tomonaga Center for the History of the Universe (TCHoU), Faculty of Pure and Applied Sciences, University of Tsukuba, \\ \hspace{0.27cm} Tsukuba, Ibaraki 305-8571, Japan \\
$^{4}$ Department of Environmental Science and Technology, Faculty of Design Technology, Osaka Sangyo University, 3-1-1, Nakagaito,\\ \hspace{0.17cm} Daito, Osaka 574-8530, Japan.\\
$^{5}$ Department of Physics, School of Advanced Science and Engineering, Faculty of Science and Engineering, Waseda University, \\ \hspace{0.17cm} 3-4-1 Okubo, Shinjuku, Tokyo 169-8555, Japan \\
$^{6}$ Waseda Research Institute for Science and Engineering, Faculty of Science and Engineering, Waseda University, 3-4-1 Okubo, \\ \hspace{0.27cm} Shinjuku, Tokyo 169-8555, Japan \\
$^{7}$ Max-Planck-Institut f\"{u}r Extraterrestrische Physik (MPE), Giessenbachstr., D-85748, Garching, Germany.\\
$^{8}$ Institute for Cosmic Ray Research, The University of Tokyo, Kashiwa, Chiba 277-8582, Japan.\\
$^{9}$ Kyoto University Research Administration Office, Yoshida-Honmachi, Sakyo-ku, Kyoto 606-8501, Japan.\\
$^{10}$ The Open University of Japan, 2-11 Wakaba, Mihama-ku, Chiba 261-8586, Japan.\\
$^{11}$ RIKEN Cluster for Pioneering Research, 2-1 Hirosawa, Wako-shi, Saitama 351-0198, Japan.\\
$^{12}$ Institute of Astronomy, Graduate School of Science, The University of Tokyo, 2-21-1, Osawa, Mitaka, Tokyo 181-0015, Japan.\\
$^{13}$ Observational Astrophysics, Department of Physics and Astronomy, Uppsala University, Box 516, SE-751 20 Uppsala, Sweden. \\
$^{14}$ Research Center for the Early Universe, Graduate School of Science, The University of Tokyo, Tokyo 113-0033, Japan.\\
$^{15}$ Department of Astronomical Science, The Graduate University for Advanced Studies (SOKENDAI), 2-21-1, Osawa, \\ \hspace{0.27cm} Mitaka, Tokyo 181-8588, Japan. \\
$^{16}$ Faculty of Science, Hokkaido University, N10 W8 Kita-ku, Sapporo 060-0810 Japan. \\
$^{17}$ Kitami Institute of Technology, 165 Koen-cho, Kitami, Hokkaido 090-8507, Japan.\\
$^{18}$ Department of Physics, Graduate School of Science, The University of Tokyo, Tokyo 113-0033, Japan.\\
$^{19}$ Kavli Institute for the Physics and Mathematics of the Universe (WPI), Todai Institutes for Advanced Study, \\ \hspace{0.27cm} The University of Tokyo, Kashiwa, Chiba 277-8583, Japan. 
}
\date{Accepted 2020 February 15. Received 2020 February 10; in original form 2020 January 8.}

\pubyear{2020}

\begin{document}
\label{firstpage}
\pagerange{\pageref{firstpage}--\pageref{lastpage}}
\maketitle

\begin{abstract}
We report on the detection of the [C\,{\sc ii}] 157.7~$\mu$m emission from the Lyman break galaxy (LBG) MACS0416\_Y1 at $z = 8.3113$, by using the Atacama Large Millimeter/submillimeter Array (ALMA). 
The luminosity ratio of [O\,{\sc iii}] 88~$\mu$m (from previous campaigns) to [C\,{\sc ii}] is 9.3 $\pm$ 2.6, indicative of hard interstellar radiation fields and/or a low covering fraction of photo-dissociation regions. 
The emission of [C\,{\sc ii}] is cospatial to the 850~$\mu$m dust emission (90~$\mu$m rest-frame, from previous campaigns), however the peak [C\,{\sc ii}] emission does not agree with the peak [O\,{\sc iii}] emission, suggesting that the lines originate from different conditions in the interstellar medium.
We fail to detect continuum emission at 1.5~mm (160~$\mu$m rest-frame) down to 18~$\mu$Jy (3$\sigma$). 
This non-detection places a strong limits on the dust-spectrum, considering the 137 $\pm$ 26~$\mu$Jy continuum emission at 850~$\mu$m. 
This suggests an unusually warm dust component ($T > 80$~K, 90\% confidence limit), and/or a steep dust-emissivity index ($\beta_\mathrm{dust} > 2$), compared to galaxy-wide dust emission found at lower redshifts (typically $T \sim 30-50$~K, $\beta_\mathrm{dust} \sim 1-2$). If such temperatures are common, this would reduce the required dust mass and relax the dust production problem at the highest redshifts. We therefore warn against the use of only single-wavelength information to derive physical properties, recommend a more thorough examination of dust temperatures in the early Universe, and stress the need for instrumentation that probes the peak of warm dust in the Epoch of Reionization.
\end{abstract}

\begin{keywords}
galaxies: formation - galaxies: high-redshift - galaxies: ISM 
\end{keywords}

%%%%%%%%%%%%%%%%%%%%%%%%%%%%%%%%%%%%%%%%%%%%%%%%%%
%%%%%%%%%%%%%%%%% BODY OF PAPER %%%%%%%%%%%%%%%%%%

\section{Introduction}
\begin{table*}
\caption{Properties of the ALMA observations}
\label{tab:ALMAlog}
\resizebox{\textwidth}{!}{\begin{tabular}{cccccc}  
\hline
\hline
UT start time & Baseline lengths & & Upper LO frequency & Integration time & PWV \\
(YYYY-MM-DD hh:mm:ss) & (m) & N$_{\textrm{ant}}$ & (GHz) & (min) & (mm) \\
\hline
2018-09-08 09:07:17 & 15 - 783 	& 43 & 203.313	& 45.7 & 1.1 \\
2018-08-30 08:50:05 & 15 - 783 	& 43 & 203.313	& 45.6 & 1.1 \\
2018-08-22 14:06:36 & 15 - 457 	& 43 & 203.313	& 45.6 & 0.6 \\ 
2018-11-25 07:42:49 & 15 - 1398 & 43 & 204.298 	& 45.1 & 0.6 \\
2018-11-25 05:20:34 & 15 - 1398 & 43 & 204.298 	& 45.1 & 0.7 \\
2018-11-25 06:32:22 & 15 - 1398 & 43 & 204.298 	& 45.1 & 0.7 \\
\hline
\end{tabular}}
\raggedright \justify \vspace{-0.2cm}
\textbf{Notes:} Reading from the left, the columns are: Column 1 - The starting time of the complete observation run; Column 2 - The distance between nearest and furthest ALMA antennae; Column 3 - The number of integrating antennae; Column 4 - The centre frequency of the upper frequency band, which contained the [C\,{\sc ii}] emission at $\sim$ 204.1 GHz; Column 5 - Total on-source integration time; Column 6 - The average precipitable water vapor during the observation run.
\end{table*}

Forming a complete picture of star-formation through cosmic time is one of the main challenges of galaxy evolution studies \citep{Madau2014}. Our current understanding of star-formation at high redshifts ($z > 7$) is mostly formed through rest-frame ultraviolet (UV) observations of Lyman-Break Galaxies (LBGs), which directly probe their stellar light (e.g. \citealt{Bouwens2015,Finkelstein2015,McLeod2015,Oesch2016,Dunlop2017,Oesch2018}).
These recent UV observations of high-redshift LBGs suggest that most star-formation is not dust-obscured, due to the short amount of cosmic time that is available for its formation, its spatial distribution across the galaxy, and its inter-stellar medium (ISM) \citep{Bouwens2016,Matthee2017}. 

The sensitivity of the Atacama Large Millimeter/submillimeter Array (ALMA) has allowed us to detect UV-selected high-redshift galaxies at submillimetre (sub-mm) wavelengths. Recent observations have revealed far-infrared spectral lines, such as bright doubly-ionized oxygen ([O\,{\sc iii}] at 88.3 $\mu$m) lines detected from LBGs out to redshifts around 8 to 9 \citep{Laporte2017,Hashimoto2018,Hashimoto2019,Tamura2019}. The [O\,{\sc iii}] line, however, only probes regions with strong UV radiation fields, close to hot O- and B-stars, which create high radiation pressures and ionized hydrogen (i.e. H\,{\sc ii} Regions).
A more complete description of the ISM requires multiple spectral line detections, each probing different phases due to their individual critical densities and temperatures. For example, the excitation energy of carbon lies slightly below the 13.6 eV required to ionize hydrogen. As such, carbon-ionizing radiation can also exist within neutral hydrogen, allowing fine-structure lines of singly-ionized carbon ($^2$P$_{3/2}$ to $^2$P$_{1/2}$ transition of C$^+$ at 157.7$\mu$m) to be emitted from neutral hydrogen regions. In the local Universe, these regions extend over larger volumes than H\,{\sc ii} regions, and combined with the prevalence of carbon, this causes the [C\,{\sc ii}]-line to be one of the most important cooling lines for the dusty ISM (e.g. \citealt{Stacey2011}). The bulk of the [C\,{\sc ii}] emission line (70 percent according to \citealt{Stacey1991,Stacey2010}) is believed to originate from photodissociation regions (PDRs), and the remainder from X-ray-dominated regions (XDRs), cosmic-ray-dominated regions (CRDRs), ionized regions (H\,{\sc ii} regions; \citealt{Meijerink2007}), low-density warm gas, shocks \citep{Appleton2013} and/or diffuse H\,{\sc i} clouds \citep{Madden1997}.

Similarly, ALMA has observed the dust continuum emission of galaxies out to redshifts of 8.38 \citep{Watson2015,Laporte2017,Tamura2019}. The amount of dust that exists at high redshifts is fundamentally constrained by the available cosmic time needed to form it. For instance, sources at $z$ = 8 have only had \textcolor{referee}{$\sim$500} million years to form all their dust and to enrich their ISM \textcolor{referee}{\citep{Tegmark1997}}. This leaves only little time to form dust in the outer shells of AGB stars, and to increase the dust mass by accretion of the ISM \citep{Ferrara2016}. Dust size growth by accretion is also hindered by strong radiation fields, whose photo-ionizing effects make the dust particles become positively charged, causing them to repel each other \citep{Ceccarelli2018}. The only known dust resource would be supernovae (SNe), although dust mass estimates at high redshift require SNe to create dust at maximum efficiency \citep{Asano2013,Gall2014,Michalowski2015,Lesniewska2019}. Potentially, stars formed by pristine gas clouds, so-called Population\,{\sc III} stars \citep{Nozawa2014}, could be an additional source of dust in the early universe both through the more traditional supernovae or during a potential red super-giant phase, however no such stars - or their tracers - have been confirmed by observations \citep{Sobral2015, Sobral2019, Bowler2017, Shibuya2018}.

Thus, providing an adequate description of the ISM conditions, whilst also probing the dust emission properties, are important goals to our understanding of galaxy evolution in the early Universe \citep{Dayal2018}.
This paper reports the detection of [C\,{\sc ii}] and the stringent upper limit on the dust continuum emission at 1.5 mm of the Lyman-break galaxy MACS0416\_Y1 at redshift 8.312. We describe the target and the observations in Section~2, and present the results in Section~3. We discuss the implications of the [C\,{\sc ii}] line strength and lack of dust continuum at 1.5~mm in Section~4, where we also compare the present result against previously observed [O\,{\sc iii}] and 850~$\mu$m dust continuum. We present our conclusions in Section~5.

Throughout this paper, we assume a flat $\Lambda$CDM cosmology with the best-fit parameters derived from the \textit{Planck} results \citep{Planck2015}, which are $\Omega_\mathrm{m} = 0.307$, $\Omega_\mathrm{\Lambda} = 0.693$ and $h = 0.693$. At $z = 8.3$, one arcsecond corresponds to a physical distance of 4.8~kpc. We also assume the \cite{Chabrier2003} initial mass function.

\section{Target and Observations}
\subsection{Target and previous observations}
\label{sec:target}
The target, MACS0416\_Y1, was initially identified as a bright $Y$-dropout galaxy at $z \approx 8$ \citep{Infante2015,Laporte2015,Laporte2016,McLeod2015,Castellano2016,Kawamata2016,GonzalezLopez2017} lying behind the MACS~J0416.1$-$2403 lensing cluster, one of the Hubble Frontier Fields \citep{Lotz2017}, with only a moderate magnification of $\mu$~=~1.43~$\pm$~0.04 \citep{Kawamata2016}.

\cite{Tamura2019} report the detection of the [O\,{\sc iii}] 88~$\mu$m line emission and 850 $\mu$m dust continuum emission. The [O\,{\sc iii}] emission confirmed the spectroscopic redshift at $z = 8.3118 \pm 0.0003$. The 850~$\mu$m dust continuum is detected at 137 $\pm$ 26 $\mu$Jy, suggesting a de-lensed total infrared luminosity of L$_{\rm{IR}}$ = (1.7 $\pm$ 0.3) $\times$ 10$^{11}$ L$_{\odot}$, assuming a modified blackbody with a dust-temperature of 50~K and an emissivity index of $\beta = 1.5$.\footnote{These results are summarised in Table~\ref{tab:photprop}.}

They further report on a UV to far-infrared spectral energy distribution modelling, where they find a moderately metal-polluted young stellar component ($\tau_{\rm{age}} = 4^{+0.7}_{-2.3}$ Myr, $Z = 0.2^{+0.16}_{-0.18}$ Z$_{\odot}$, M$_{*}$ = 2.4$^{+0.7}_{-0.1} \times 10^8$ M$_{\odot}$, assuming the Calzetti extinction law), with a star-formation rate (SFR) of around 60 M$_{\odot}$/yr. However, an analytic dust mass evolution model assuming only a single episode of star-formation does not reproduce the metallicity and dust mass within these 4 Myr, suggesting a pre-existing evolved stellar component with a M$_{\rm{star}}$ $\sim$ 3 $\times$ 10$^9$ M$_{\odot}$ and an age $\tau_{\rm{age}}$ $\sim$ 300 Myr, around half the age of the Universe at $z = 8.312$. 

\subsection{ALMA observations}
\label{sec:observations}
The ALMA observations were initially carried out in August, September and November 2018 as a Cycle 5 program (Program ID: 2017.1.00225.S, P.I. Y. Tamura). We summarise the observations in Table \ref{tab:ALMAlog}. These observations were executed in the C43-1, -2, -3 and -4 configurations, with baseline lengths ranging from 15 to 1400 meters. The precipitable water vapour (PWV) ranges from 0.6 to 1.2~mm.
The local oscillator frequency of the Band 5 receivers \textcolor{referee}{\citep{Belitsky2018}} was set from 200.5~GHz to 204.85~GHz, in order to cover the [C\,{\sc ii}] 158~$\mu$m line at $z$~=~8.31, which was expected at 204.10~GHz. 
In total, 272.2 minutes were spent effectively on-source. 
The phase-tracking center was set to the UV continuum position ($\alpha_{\rm{ICRS}}$, $\delta_{\rm{ICRS}}$) = (04$^{\rm{h}}$16$^{\rm{m}}$09\fs 4010, $-24^{\circ}$16$^{{\rm{'}}}$35\farcs 470). Two quasars, J0348$-$2749 and J0453$-$2807, were used for complex gain calibration, and another quasar, J0522$-$3627, was used for bandpass calibration.

The calibration and flagging were done using the standard pipeline running on CASA \citep{McMullin2007} version 5.4.1. A CASA task \texttt{tclean} was used to \textcolor{referee}{combine all visibilities into an} image, both for the continuum and the [C\,{\sc ii}] emission, where the [C\,{\sc ii}]-containing channels were omitted in the continuum image. The data cube was sampled at 35~km~/s ($\sim$23.3 MHz) to search for the [C\,{\sc ii}] line feature. We used the natural weighting on the \texttt{tclean} task to maximize the point-source flux significance. We integrate the data cube between $-140$ and +140 km/s around the [C\,{\sc ii}] frequency to generate the [C\,{\sc ii}]-integrated map\footnote{Integrating the data cube between larger velocity widths does not result in a significant increase in the integrated flux.}, resulting in a beam-size of 0$\farcs$46 $\times$ 0$\farcs$64 at a position angle of $-$77 degrees, and an r.m.s. noise of 9 mJy/beam. For the continuum map, the resulting beam-size is 0$\farcs$59 $\times$ 0$\farcs$78 at a position angle of $-$81 degrees, and r.m.s. noise is 6 $\mu$Jy/beam. This $-140$ and $+$140 km/s velocity range was chosen because it resulted in the highest signal-to-noise ratio for the [C\,{\sc ii}] emission, and agrees with a by-eye inspection of the spectrum. \textcolor{referee}{Here, we calculate the r.m.s. noise from the off-source standard deviation of the maps.}

\section{Results}
\label{sec:results}
\begin{figure*}
\includegraphics[width=\linewidth]{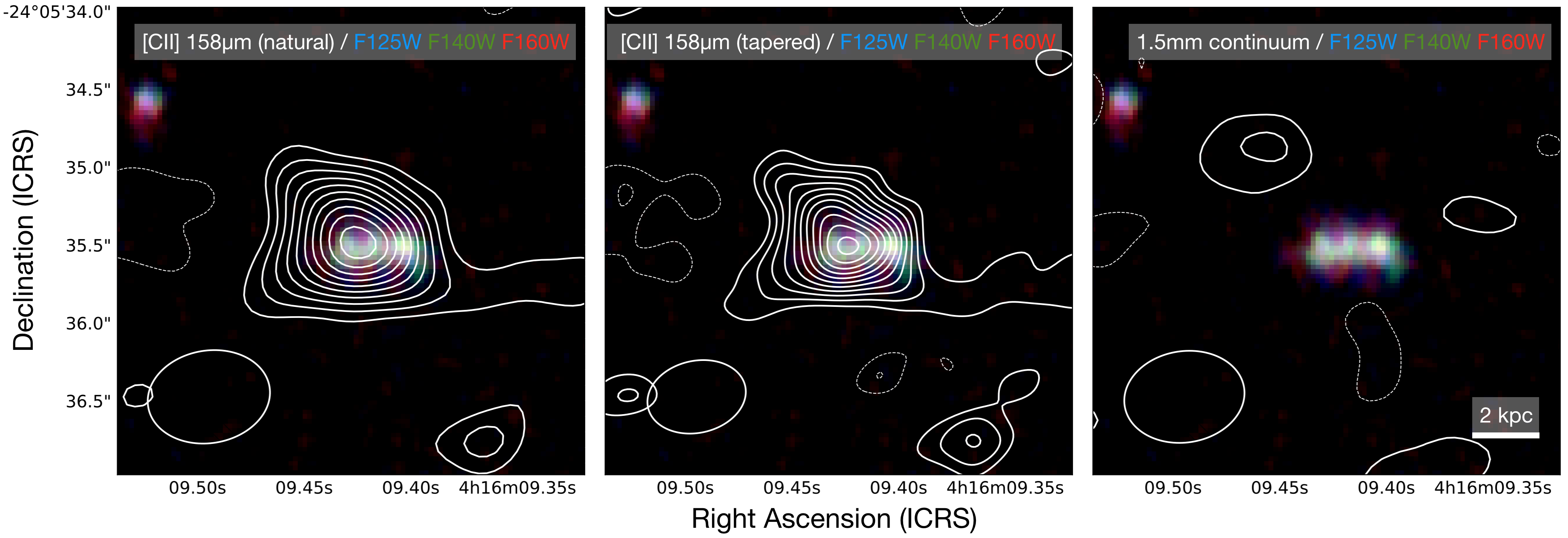}
\caption{(\textit{Left and Middle}) The ALMA-observed [C\,{\sc ii}] 158~$\mu$m emission of MACS0416\_Y1 (-140 and +140 km/s around the [C\,{\sc ii}] frequency), naturally-weighted (\textit{left}) and tapered by a Gaussian kernel with FWHM = 0$\farcs$27 (\textit{middle}), shown as contours drawn at $-$2, 2, 3, ..., 11 and 12 $\sigma$, respectively, where the negative contours are drawn as dashed lines. (\textit{Right}) The non-detected, tapered (by a Gaussian kernel with FWHM = 0$\farcs$27) dust continuum image at a rest-frame wavelength of 160~$\mu$m, shown as contours drawn at $-$2, 2, 3 $\sigma$, where negative contours are indicated with dashed lines. The bottom-left ellipse in each panel shows the beam-size for the contours. All backgrounds are the HST/WFC3 near-infrared false-colour RGB image is composed of F160W (R), F140W (G) and F125W (B). The Hubble imaging suggests MACS0416\_Y1 consists of eastern, central and western components in rest-frame UV emission.}
\label{fig:HubbleRGB}
\end{figure*}
\begin{figure}
\includegraphics[width=\linewidth]{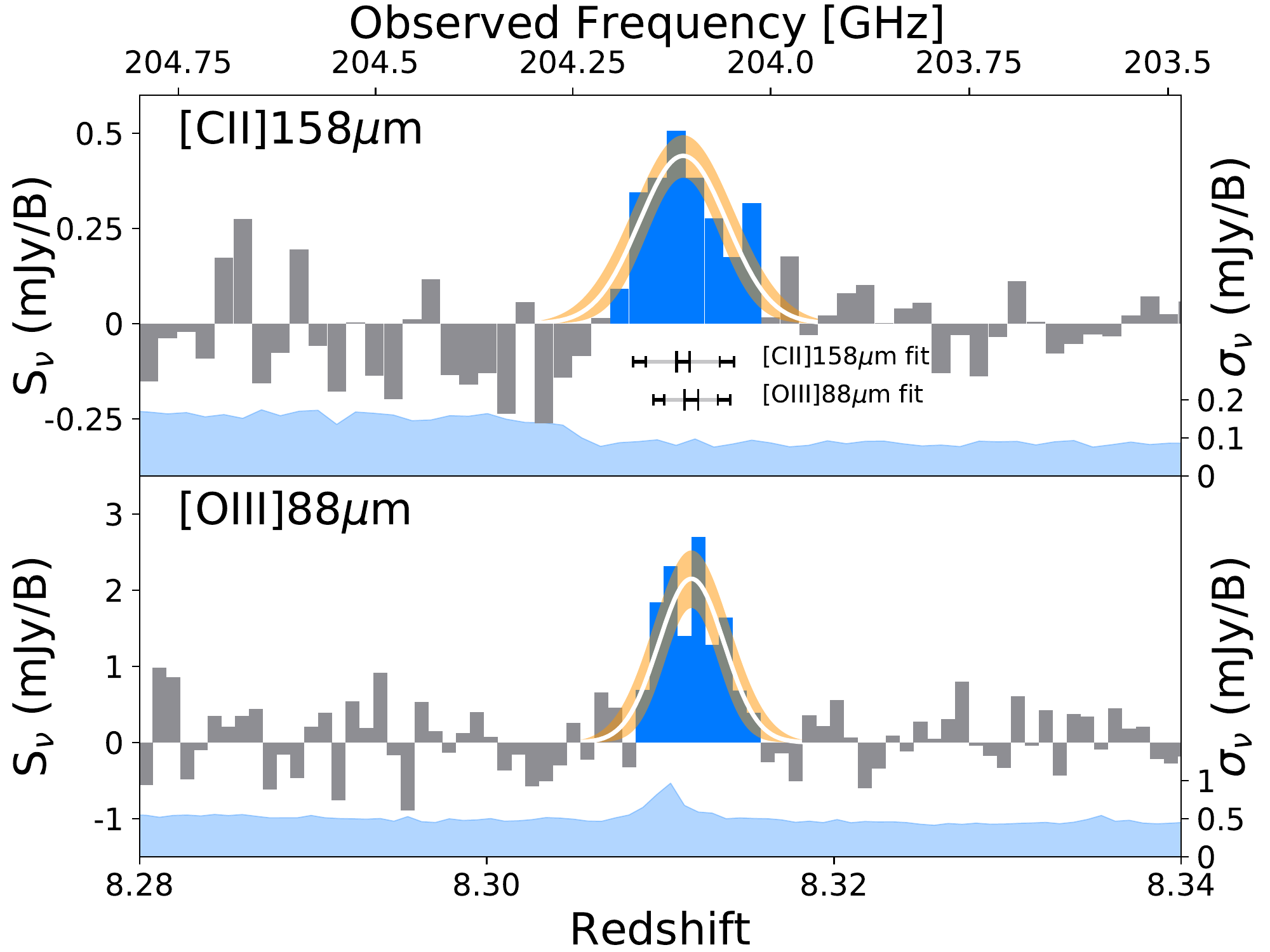}
\caption{(\textit{Top}) The ALMA-spectrum of MACS0416\_Y1 showing the [C\,{\sc ii}] emission, binned at 35 km/s, extracted at the peak of the [C\,{\sc ii}] emission. The orange line shows the 16th to 84th percentile Gaussian fit of the spectral line. The blue bins indicate those stacked to form the [C\,{\sc ii}] image. The light-blue filled line in the bottom of the panel indicates the per-bin noise level ($\sigma_{\nu}$), with the scale shown on the right-hand side. The horizontal lines underneath the [C\,{\sc ii}] line profile show the 16th to 84th percentile spread of the redshifts and line widths of both [C\,{\sc ii}] and [O\,{\sc iii}] for comparison
(\textit{Bottom}) Similar to the top figure, we show the continuum-subtracted [O\,{\sc iii}] ALMA spectrum for comparison to the [C\,{\sc ii}] emission, from \citet{Tamura2019}, and extracted at the peak of the [O\,{\sc iii}] emission.}
\label{fig:Linefit35.pdf} 
\end{figure}

\subsection{Detection of the [C\texorpdfstring{\,{\sc ii}}{II}] 158 micron emission}
We detect the [C\,{\sc ii}] 158~$\mu$m emission at the position of MACS0416\_Y1. Figure \ref{fig:HubbleRGB} shows the Hubble false-colour RGB image, overplotted with the contours of a naturally-weighted [C\,{\sc ii}] map (\textit{left}), a tapered [C\,{\sc ii}] (\textit{middle}), and the [C\,{\sc ii}]-subtracted tapered continuum map (\textit{right}). We create the tapered maps by convolving our naturally-weighted maps using a Gaussian kernel with a full-width at half maximum (FWHM) of 0$\farcs$27. We find this FWHM size by maximizing the signal-to-noise of the [C\,{\sc ii}] emission for all tapering sizes between 0$\farcs$01 and 0$\farcs$50, \textcolor{referee}{which covers the potential angular sizes of MACS0416\_Y1 expected from the [OIII] emission}. As noted in \cite{Tamura2019}, the Hubble image appears to be a multi-component system, with eastern, central and western components. They note that the 90~$\mu$m rest-frame dust continuum is centered on the eastern component, whilst the [O\,{\sc iii}]~88$\mu$m emission appears to be more centrally-located.\footnote{The astrometry of the Hubble images are set using GAIA stars.}  

We calculate the central position and total apparent flux by fitting a 2D Gaussian to the naturally-weighted integrated intensity image, since it has the best positional accuracy. The central position of the [C\,{\sc ii}] emission is at RA$_{\rm ICRS}$: 04$^{\rm{h}}$16$^{\rm{m}}$09\fs426 and DEC$_{\rm ICRS}$: $-24^{\circ}$05$^{{\rm{'}}}$35\farcs470 with a positional uncertainty of $(\sigma_\mathrm{RA},\,\sigma_\mathrm{Dec}) = (0\farcs033,\,0\farcs026)$. This position lies between the eastern and central component of MACS0416\_Y1, which was also traced by the dust-component in \cite{Tamura2019}. Deconvolution of the [C\,{\sc ii}]-emission reveals a barely-resolved source-size of 0.48 $\pm$ 0.14 by 0.16 $\pm$ 0.25 arcsecond, at a position angle of 36 $\pm$ 36 degrees, which agrees with the angular size found in the [O\,{\sc iii}] observations, whilst non-axisymmetric spur-like structure is apparent in the image. This corresponds to a physical size of 2.3 $\pm$ 0.7 by 0.8 $\pm$ 1.2 kpc at $z = 8.3$. 

We use this 2D Gaussian fit to measure an apparent [C\,{\sc ii}] flux of 130.2 $\pm$ 20.4 mJy km/s on the naturally-weighted image. In order to maximise the signal-to-noise ratio of the spectrum and fitted line-profile, we taper the [C\,{\sc ii}] map with a 0.27 arcsecond tapering size. The spectrum and a Gaussian fit to the spectrum are shown in Figure~\ref{fig:Linefit35.pdf}, where the top panel shows the [C\,{\sc ii}] emission, and the bottom panel shows the [O\,{\sc iii}]~88$\mu$m emission from \cite{Tamura2019} in order to compare the line profiles. The filled area at the bottom of each panel represents the uncertainty $\sigma_{\nu}$ in each frequency bin, whose axes are shown on the right-hand side of the figure. The two horizontal lines below the [C\,{\sc ii}] line indicate the fitted average frequency and line width of both the [C\,{\sc ii}] and [O\,{\sc iii}] lines. 

The best-fit Gaussian profile finds a central frequency at 204.1105 $\pm$ 0.0082 GHz, indicating a redshift of 8.31132 $\pm$ 0.00037. This central frequency of the [C\,{\sc ii}]-line suggests a velocity offset from the [O\,{\sc iii}]-line of $\sim$15 $\pm$ 15\,km/s. 
The combined [O\,{\sc iii}] and [C\,{\sc ii}] estimated redshift is $\rm \bar{z}$ = 8.31161 $\pm$ 0.00023, calculated using
\begin{equation}
\rm \bar{z} = \sqrt{\frac{\sum^i{\left( z_i/dz_i \right)^2}}{\sum^i \left( 1/dz_i \right)^2}}.
\label{eq:specz}
\end{equation} 
Here, $\rm z_i$ is the redshift of line i, and $\rm dz_i$ is the uncertainty in line~i. 
The [C\,{\sc ii}] line width (FWHM) is 191 $\pm$ 29 km/s, which is slightly larger than the line width measured for [O\,{\sc iii}] (141 $\pm$ 21 km/s), with a $\Delta$FWHM of 50 $\pm$ 36\,km/s, and could be due to the difference in regions where [O\,{\sc iii}] and [C\,{\sc ii}] are predominantly emitted. We derive the line luminosity using
\begin{equation}
    L_{\rm{line}} = 1.04 \times 10^{-3} \times \left(\frac{S_{\rm{line}} \Delta v}{\textrm{Jy km/s}}\right) \left(\frac{D_{\rm{L}}}{\rm{Mpc}}\right)^2\left(\frac{\nu_{\rm{obs}}}{\rm{GHz}}\right)L_{\odot},
\end{equation}
from \cite{Solomon2005}, where $S_{\rm{line}}$ is the velocity-integrated flux corrected for lensing, $D_{\rm{L}}$ the luminosity distance, and $\nu_{\rm{obs}}$ the observed line frequency. We find (1.40 $\pm$ 0.22) $\times$ $10^8$ $L_{\odot}$ for the magnification-corrected [C\,{\sc ii}] line luminosity. We do not account for the effect of the Cosmic Microwave Background (CMB) on the [C\,{\sc ii}]-luminosity, as this is dependent on the properties of the [C\,{\sc ii}]-emitting regions. The high dust temperature we find in Section \ref{sec:15mmnondetection} could suggest that the regions where [C\,{\sc ii}] is emitted are also warm, significantly decreasing the CMB attenuation \citep{Lagache2018}. Moreover, the effect of the CMB on the [C\,{\sc ii}] emission decreases at redshifts greater than 7, suggesting the effect of the CMB is minor. \cite{Laporte2019} reported on the non-detection of [C\,{\sc ii}] from a LBG with a similar redshift ($z$ = 8.38), though with only half the star-formation rate of MACS0416\_Y1, and suggested the [C\,{\sc ii}] luminosity could be under-estimated by a factor of $\sim$3 in the most extreme case of CMB attenuation using \cite{Lagache2018}.

\subsection{Upper-limit of 1.5 mm Dust Continuum}
We did not detect the dust continuum emission at 1.5~mm (200~GHz) in either the tapered or naturally-weighted map, after we combine all the spectral windows, excluding the channels that contain the [C\,{\sc ii}]-emission.
We find a 3$\sigma$ upper limit of 18~$\mu$Jy at 1.5~mm (160~$\mu$m rest-frame). This is significantly lower than we expected from the 137 $\pm$ 26 $\mu$Jy dust continuum detected at 850~$\mu$m (90~$\mu$m rest-frame). In the case of a modified blackbody with a temperature of  T$_{\rm{dust}}$ = 50 K and a dust-emissivity of $\beta$ = 1.5, we would expect a 1.5~mm flux of around 36 $\mu$Jy, well above the detection limit ($\sim6\sigma$). We more thoroughly discuss the consequences of this non-detection in Section \ref{sec:dis_cont}.

\begin{table}
\caption{Line and continuum properties of MACS0416\_Y1}
\label{tab:photprop}
\resizebox{0.8\linewidth}{!}{\begin{tabular}{ll}  
\hline \hline
Parameter & Value \\
\hline \multicolumn{2}{c}{\textbf{Our observations}} \\
\hline
$S_{\rm{1.5 mm}}$$^{\ddagger}$ 		& < 18 $\mu$Jy (3$\sigma$) 			\\
FWHM$_{\rm{1.5 mm}}$$^{\ddagger}$ 	& 0\farcs 59 $\times$ 0\farcs 78			\\
$F_{\rm{[C\,{\sc II}]}}$$^{\ddagger}$ 		& 130.2 $\pm$ 20.4 mJy km/s		\\
$\Delta$v 				& 191 $\pm$ 29 km/s		\\
$z_{\rm{[C\,{\sc II}]}}$		& 8.31132 $\pm$ 0.00037 			\\
$L_{\rm{[C\,{\sc II}]}}$	    & (1.40 $\pm$ 0.22) $\times$ 10$^8$ $L_{\odot}$ \\
\hline 
\multicolumn{2}{c}{\textbf{Previous ALMA observations}$^{\dagger}$} \\
\hline
$S_{\rm{850\mu m}}$$^{\ddagger}$ 	& 137 $\pm$ 26 $\mu$Jy 			\\
FWHM$_{\rm{850\mu m}}$$^{\ddagger}$ 	& $0\farcs 36 \times 0\farcs 10$			\\
$S_{\rm{1140\mu m}}$$^{\ddagger}$ 	& $< 116$ (2$\sigma$) $\mu$Jy \\
$F_{\rm{[O\,{\sc III}]}}$$^{\ddagger}$ 		& 660 $\pm$ 160 mJy km/s		\\
$\Delta$v				& 141 $\pm$ 21 km/s		\\
$z_{\rm{[O\,{\sc III}]}}$		& 8.3118 $\pm$ 0.0003 			\\
$L_{\rm{[O\,{\sc III}]}}$		& (1.2 $\pm$ 0.3) $\times$ 10$^9$ $L_{\odot}$ \\ \hline
\end{tabular}}
\raggedright \justify \vspace{-0.2cm}
$^{\dagger}$ Values are from \cite{Tamura2019}.\\
\textcolor{referee}{$^{\ddagger}$ Not corrected for lensing magnification}
\end{table}

\section{Discussion}
\subsection{[O\texorpdfstring{\,{\sc iii}}{III}]-to-[C\texorpdfstring{\,{\sc ii}}{II}] luminosity ratio}
\label{sec:OIIIvsCIILumRatio}
\begin{figure}
\includegraphics[width=\linewidth]{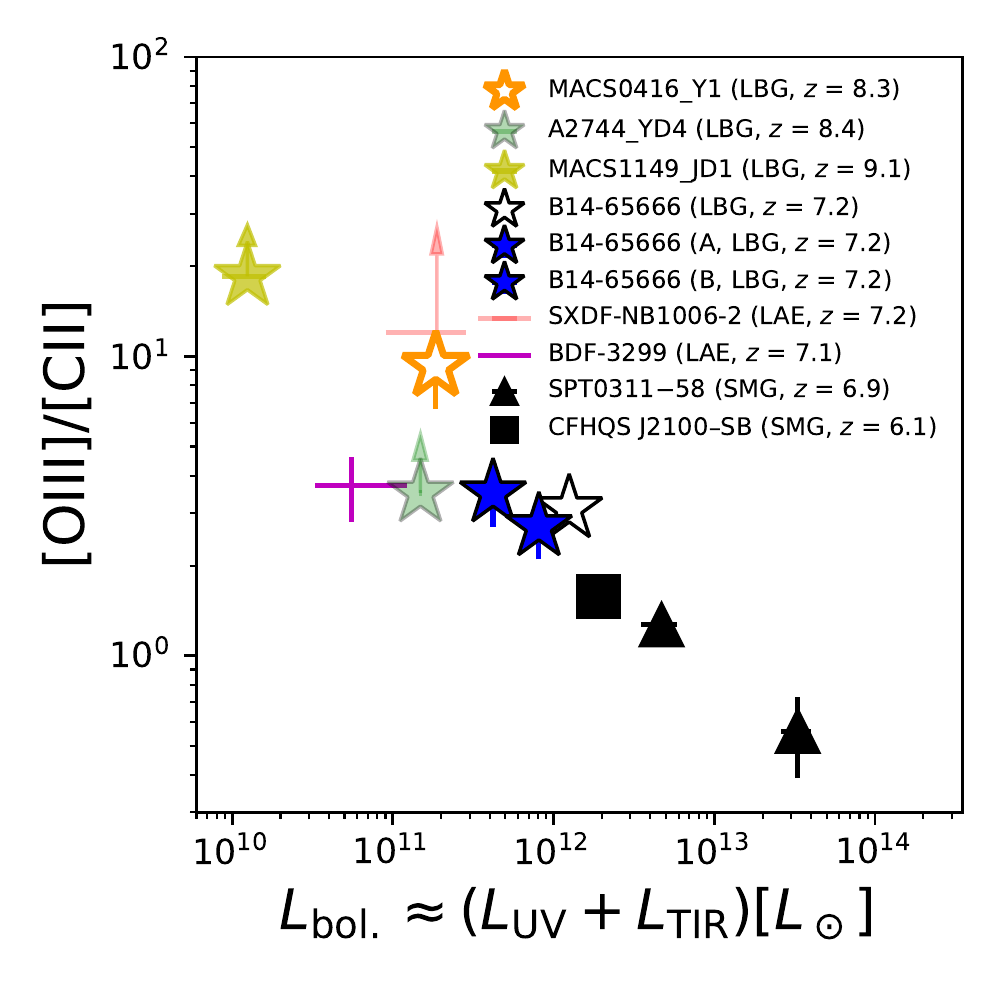}
\caption{The [O\,{\sc iii}]-to-[C\,{\sc ii}] luminosity ratio against the total bolometric luminosity, a combination of both UV and FIR components. We find that all high-redshift LBG and LAE have a ratio larger than unity, and note a power-law trend. We describe the sources further in the text.}
\label{fig:Hashimoto}
\end{figure}
We compare the oxygen to carbon luminosity ratio to the total bolometric luminosity in Figure \ref{fig:Hashimoto}. Here, the [O\,{\sc iii}] luminosity is from \cite{Tamura2019}, who finds (1.2 $\pm$ 0.3) $\times$ 10$^9$ L$_{\odot}$, resulting in a [O\,{\sc iii}]-to-[C\,{\sc ii}] luminosity ratio of 9.3 $\pm$ 2.6, not taking CMB-dimming into account. The delensed UV luminosity of MACS0416\_Y1 is 4.5 $\times$ 10$^{10}$ L$_{\odot}$, taken from \cite{Laporte2015}.
The far-infrared luminosity is 1.7 $\times$ 10$^{11}$ L$_{\odot}$, assuming T$_{\rm{dust}}$ = 50~K and a dust emissivity factor ($\beta$) of 1.5. While we show in Section\,\ref{sec:15mmnondetection} that this temperature is most likely higher, for fair comparison, we use this temperature for our far-infrared luminosity calculation, since all other sources have assumed dust temperatures in the range of 30 to 60~K.

We compare our galaxy against other high-redshift LBGs, Lyman-$\alpha$ Emitters (LAEs) and sub-mm galaxies (SMGs) that have observations targeting both the [O\,{\sc iii}] and [C\,{\sc ii}] lines. Firstly, we use the [C\,{\sc ii}] upper limits, combined with the far-infrared luminosities and [O\,{\sc iii}] of the $z$ = 8.38 and 9.11 LBGs A2744\_YD4 and MACS1149\_JD1 \citep{Laporte2019}. We show the $z$ = 7.15 LBG B14-65666 as a single luminosity-averaged source, as well as its two resolved components, from \cite{Hashimoto2019}. We show the upper limit for the $z$ = 7.2 LAE SXDF-NB1006-2 described in \cite{Inoue2016}, and the $z$ = 7.1 LAE BDF-3299 detailed in \cite{Carniani2017}. Finally, we also show three $z$ $>$ 6 SMGs, the eastern and western components of SPT0311-58, and CFHQS J2100-SB, described in \cite{Marrone2018} and \cite{Walter2018} respectively, in order to compare with more dusty and star-forming galaxies.
We find a high [O\,{\sc iii}]-to-[C\,{\sc ii}] ratio for MACS0416\_Y1, compared to other sources with detected [C\,{\sc ii}] emission. \textcolor{referee}{We note that the downward trend in [O\,{\sc iii}]-to-[C\,{\sc ii}] for increasing bolometric luminosities, as mentioned in \cite{Hashimoto2019}, appears consistent also for fainter sources (e.g. MACS1149\_JD1).}

Local dwarf galaxy studies \citep{Cormier2015,Cormier2019} find [O\,{\sc iii}]-to-[C\,{\sc ii}] luminosity ratios between 2 and 10 for these 0.02 to 1~Z$_{\odot}$ galaxies. The low metallicity of high-redshift galaxies could thus be contributing to the high ratio, where the low metallicity allows hard radiation fields to extend over longer distances, which is necessary to produce [O\,{\sc iii}] emission. The porosity of the neutral media in these high-redshift galaxies such as MACS0416\_Y1 could also play a role, since the modeling of dwarf galaxies by \cite{Cormier2019} has showed that almost all dwarf galaxies have H\,{\sc ii} regions that are directly exposed to the IGM, without shielding by a PDR.
Simulations by \cite{Arata2019} suggest this porosity could be due to the shallow gravitational potential in the recently-formed dark-matter haloes of high-redshift LBGs such as MACS0416\_Y1. Their simulations combine cosmological zoom-in hydrodynamic simulations and radiative transfer modeling to investigate three massive galaxies (and their satellites). The total halo masses of these galaxies range between 2.4 to 70~$\times{}$~$10^{10}$~M$_{\odot}$. For reference, assuming a baryonic mass of 1~$\times{}$~$10^{10}$~M$_{\odot}$ \citep{Tamura2019}, and a baryonic-to-dark matter ratio of 5-to-27, the halo mass of MACS0416\_Y1 is expected to be $\sim$6~$\times{}$~$10^{10}$~M$_{\odot}$.
They showed that this shallower gravitational potential allows stellar feedback to regulate star-formation, alternating between a galaxy in an infrared-luminous and a UV-luminous phase. During each phase, the spectrum is dominated either by emission from dust-obscured star-formation inside of PDRs during the infrared-luminous phase or by unobscured emission from exposed H\,{\sc ii} regions with little dust obscuration during the UV-luminous phase. Each phase is characterized by a specific [O\,{\sc iii}]-to-[C\,{\sc ii}] ratio, where the dust-obscured phase has strong [O\,{\sc iii}] emission, while the [O\,{\sc iii}]-emission becomes faint rapidly during the UV-luminous phase \citep{Arata2020}. %Alternatively, several other studies find surpression of [C\,{\sc ii}] emission in strong radiation fields in dusty star-forming galaxies (e.g. \citealt{Rybak2019}). 

\cite{Pallottini2019}, and previously \cite{Pallottini2017_Structure} and \cite{Vallini2017}, have modelled the formation and evolution of several LBGs in the early Universe using cosmological zoom-in simulations, accounting for ISM evolution and using radiative transfer modelling. Their one-dimensional radiative transfer modelling showed that the efficiency of [C\,{\sc ii}] emission (luminosity over mass) drops for extreme PDR densities greater than 10$^4$ cm$^{-3}$. [C\,{\sc ii}] emission from MACS0416\_Y1 could also be suppressed by the photo-electric effect of UV photons on the dust, creating free electrons that remove ionized carbon from the ISM by recombination into neutral carbon. These findings are thus consistent with strong interstellar radiation fields, given both the relatively-low [C\,{\sc ii}] emission from MACS0416\_Y1, and with its warm dust temperature, as we will discuss in Section \ref{sec:15mmnondetection}.

Simulations are, however, unable to accurately recreate the global [O\,{\sc iii}]-to-[C\,{\sc ii}] ratios seen in MACS0416\_Y1 and other high-redshift galaxies. In \cite{Pallottini2019}, small regions of their simulated galaxies show [O\,{\sc iii}]-to-[C\,{\sc ii}] ratios of around 10, but the global ratio of each galaxy is around 0.3. In their simulations, the most massive galaxy has a stellar mass of 4.2 $\times$ 10$^9$~M$_{\odot}$. They do find that [C\,{\sc ii}] is suppressed by the starburst-phase of the galaxy, disrupting and photo-dissociating the emitting molecular clouds around star-formation sites. The simulations by \cite{Katz2019} find only one galaxy with a [O\,{\sc iii}]-to-[C\,{\sc ii}] ratio greater than unity, although its star-formation rate is less than 1~M$_{\odot}$/yr, and most galaxies with SFRs > 10~M$_{\odot}$/yr have ratios ranging from 0.1 down to 0.02. Their simulations are based on a suite of cosmological radiation-hydrodynamics simulations centred on LBGs that follow from adaptive-mesh cosmological simulations assuming the \cite{Planck2015} cosmology. These cosmological simulations evolve the inhomogeneities in temperature, density, metallicity, ionisation parameter and spectral hardness. Their three most massive sources range between 4 to 42 $\times$ 10$^9$~M$_{\odot}$ at $z$ = 9. 

\cite{Harikane2019} also finds a high [O\,{\sc iii}]-to-[C\,{\sc ii}] ratio for their sources, and they model the origins of this high ratio using the CLOUDY photo-ionization code \citep{Ferland2017}. Their detailed discussion of the potential causes of their high line ratio addresses (A) a higher ionization parameter, (B) a lower gas metallicity, (C) a higher gas density, (D) a lower C/O abundance ratio, (E) a lower PDR covering fraction, (F) CMB attenuation effects, (G) a spatially extended [C\,{\sc ii}] halo, and (H) inclination effects. Their modelling suggests that the high [O\,{\sc iii}]-to-[C\,{\sc ii}] ratio of their sources, and of MACS0416\_Y1, can be explained by either the higher ionization parameter (A) or the lower PDR covering fraction (E). Similarly, a combination of a higher gas density (C), a lower C/O abundance ratio (D) and the effect of the CMB (F) are adequate for explaining the high ratio of MACS0416\_Y1. Their CLOUDY modelling is able to exclude the low metallicity (B) as an explanation for the high ratio of their sources and of MACS0416\_Y1.
The [C\,{\sc ii}] flux of MACS0416\_Y1 does not appear to increase for larger apertures, disqualifying missed [C\,{\sc ii}] emission due to a spatially extended [C\,{\sc ii}] halo (G) around MACS0416\_Y1. Finally, the inclination effects (H) can be disqualified since this should affect both the [C\,{\sc ii}] and [O\,{\sc iii}] lines of MACS0416\_Y1 to a similar extent, especially since the line profiles (Figure \ref{fig:Linefit35.pdf}) suggest they lie in the same plane.

\subsection{[C\texorpdfstring{\,{\sc ii}}{II}] line luminosity, profile and geometry}
\subsubsection*{Line luminosity}
\begin{figure}
\includegraphics[width=\linewidth]{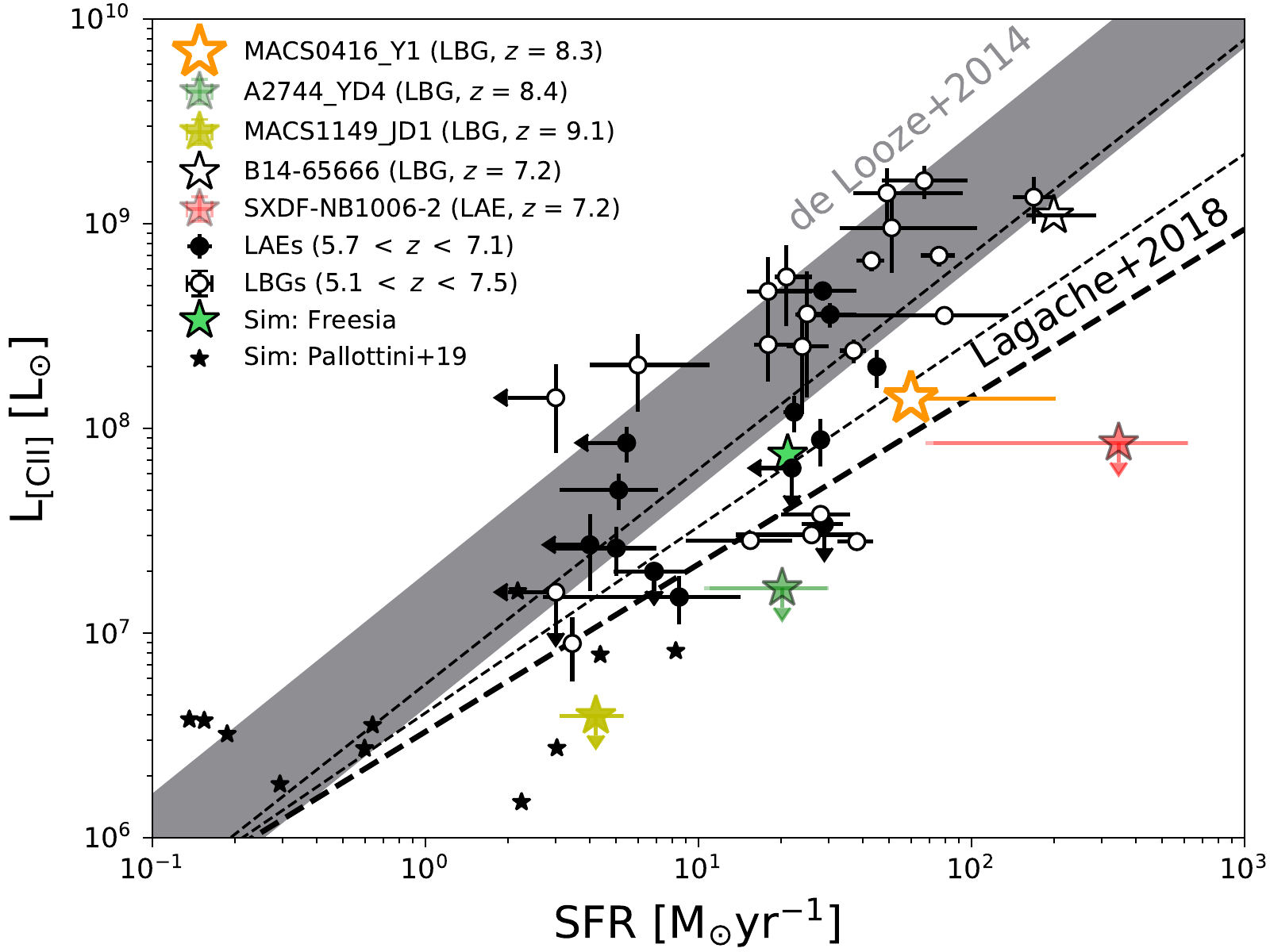}
\caption{The [C\,{\sc ii}] line luminosity compared to the total (UV + IR) star-formation rate for MACS0416\_Y1, high redshift sources, simulations (\citet{Pallottini2019}, separately indicating the brightest galaxy: Freesia) and the scaling relations from \citet{deLooze2014} and \citet{Lagache2018}. \textcolor{referee}{We show the \citet{Lagache2018} scaling relations for $z = 8.3$ (\textit{bold dashed line}), and $z=5$ and $=7$ (\textit{thin dashed lines}).} We find a slight [C\,{\sc ii}] deficit among the highest-redshift sources, when compared to other LBGs and the relation from \citet{deLooze2014}. We find good agreement with the scaling relation from \citet{Lagache2018}.}
\label{fig:deLooze}
\end{figure}
We plot the [C\,{\sc ii}] line luminosity against the integrated (UV + IR) star-formation rate (SFR) in Figure \ref{fig:deLooze}, where we take the SFR for MACS0416\_Y1 from \cite{Tamura2019}. They assume a Chabrier IMF to calculate the total star-formation rate \citep{Chabrier2003} using UV-to-FIR spectral modelling tool PANHIT\footnote{Panchromatic Analysis for Nature of HIgh-z galaxies Tool (PANHIT), \url{http://www.icrr.u-tokyo.ac.jp/\~mawatari/PANHIT/ PANHIT.html}} (\citealt{Mawatari2016}, \citealt{Hashimoto2018}, and \citealt{Mawatari2020}).
We compare MACS0416\_Y1 against high-redshift observations, simulations from \cite{Pallottini2019} and two [C\,{\sc ii}]-SFR$_{\rm{FIR}}$ scaling relations from \cite{deLooze2014} and \cite{Lagache2018}. We show the [C\,{\sc ii}] upper-limits and SFR-estimates of the $z$ = 8.38 and 9.11 LBGs A2744\_YD4 and MACS1149\_JD1 from \cite{Laporte2019}, where the star-formation rate of the $z$ = 8.38 galaxy is calculated by MAGPHYS and of the $z$ = 9.11 is calculated using PANHIT model \citep{Hashimoto2018}. We show the $z$ = 7.15 LBG B14-65666 from \cite{Hashimoto2019}, where the SFR is calculated using the same SED fitting model as MACS0416\_Y1. We also show the upper limit for the $z$ = 7.2 LAE SXDF-NB1006-2 described in \cite{Inoue2016}, and use the star-formation rate calculated using a Chabrier IMF, taken from \cite{Harikane2019}. The high-redshift observations refer to a combination of 5.7 $<$ $z$ $<$ 7.1 LAEs from \cite{Bradac2017,Carniani2017,Carniani2018,Kanekar2013,Matthee2017,Ota2014,Smit2018} and 5.1 $<$ $z$ $<$ 7.5 LBGs from \cite{Barisic2017,Capak2015,Jones2017,Knudsen2016,Schaerer2015,Willott2015}.
We compare against simulations by \cite{Pallottini2019}, highlighting its most massive galaxy (Freesia, M$_{\rm star}$ = 4.2 $\times$ 10$^9$~M$_{\odot}$, \citealt{Pallottini2017_Structure,Pallottini2017_Chemistry}). The \cite{deLooze2014} scaling relation is based on the star-formation rate derived from UV and mid-IR observations of both low-metallicity local dwarf galaxies. The \cite{Lagache2018} scaling relation is based on the combination of a semi-analytical model (G.A.S., \citealt{Cousin2019}) and the CLOUDY photo-ionization code \citep{Ferland2017}, where they assume PDRs are the dominant region where the [C\,{\sc ii}] is emitted. Their redshift-dependent [C\,{\sc ii}] scaling relation agrees with recently-observed, high redshift observations of UV- and blindly-selected sources, although the scaling relation has difficulty in reproducing the [C\,{\sc ii}] emission from high-z sub-mm galaxies. Here, we show the scaling relation for $z = 8.3$ \textcolor{referee}{(\textit{bold dashed line}) and for $z=5$ to $z=7$ (\textit{thin dashed lines})}. For lower redshifts, this line steepens and rises.

The [C\,{\sc ii}] emission of our $z = 8.31$ LBG, MACS0416\_Y1, appears fainter than predicted by the relation of \cite{deLooze2014}. This is also seen for the other $z>8$ galaxies, which all appear to lie below the scaling relation from \cite{deLooze2014}. The high-redshift LAEs and the simulated LBGs in \cite{Pallottini2019} also exhibit a deficit in the [C\,{\sc ii}] emission, while this is not seen for actual observations of $z<8$ LBGs in general. 
The scaling relation by \cite{Lagache2018} agrees with our [C\,{\sc ii}] luminosity, and appears to agree most of the $z>8$ LBGs and the LAEs. \textcolor{referee}{The relationship seems to under-predict the [CII] luminosity of the 5.1 $<$ z $<$ 7.5 LBGs (\textit{black-and-white circles}), as well as for $z = 7.2$ LBG B14-65666 (\textit{black star}, \citealt{Hashimoto2019}).}

The LAEs typically lie below the LBGs in Figure \ref{fig:deLooze}, except for the $z>8$ LBGs, such as MACS0416\_Y1. Lyman-$\alpha$ emission originates in hard radiation fields, which deplete the neutral medium, and therefore decrease the [C\,{\sc ii}]-luminosity \citep{Cormier2015,Cormier2019}. For increasing redshift, however, it becomes more difficult to detect this Ly-$\alpha$ emission since the Universe is not yet ionized at those redshifts ($z_{\rm re}$ = 7.7 $\pm$ 0.8, \citealt{PlanckReioniziation2018}), and at $z=8.3$, MACS0416\_Y1 is likely located in a mostly-neutral inter-galactic medium. $z>8$ LBGs such as MACS0416\_Y1 could thus more closely resemble lower redshift LAEs than LBGs.

The position of MACS0416\_Y1, compared to the \cite{deLooze2014} scaling relation, agrees with the simulations by \cite{Pallottini2019}. Their spatially-resolved simulations indicate that the larger molecular clouds shield the [C\,{\sc ii}] efficiently, decreasing the [C\,{\sc ii}] emission of these regions to around an order of magnitude below the expected behaviour from \cite{deLooze2014}. Instead, the \cite{Lagache2018} scaling relation seems to agree with these simulations as well as MACS0416\_Y1, and also reproduces the flatter slope in the [C\,{\sc ii}]-SFR scaling relation seen in these simulations.
The lack of [C\,{\sc ii}] emission seen for the $z$ = 8.38 and $z$ = 9.11 LBGs \citep{Laporte2019} could be due to the large scatter on the scaling relations ($\sim$ 0.5 dex), or potentially due to inclination effects, where the added velocity-width of an edge-on galaxy pushes the line peak below the detection limit (e.g. \citealt{Kohandel2019}). CLOUDY modeling by \cite{Ferrara2019} attributes the lack of [C\,{\sc ii}] luminosity compared to the local \cite{deLooze2014} scaling relation for MACS0416\_Y1 and other $z > 8$ galaxies to three factors, namely (i) bursty star-formation, causing high SFRs for their [C\,{\sc ii}] luminosity, (ii) low metallicity, and (iii) low gas density. Only the first factor, bursty star-formation, applies to MACS0416\_Y1, as both its star-formation rate and gas density are too high to cause a decrease in [C\,{\sc ii}] luminosity. Bursty episodes of star-formation are also seen in modeling by \cite{Arata2019}, and this burstiness produces a similar scatter on the [C\,{\sc ii}]-SFR diagram as predicted by \cite{Ferrara2019} \citep{Arata2020}.

\subsubsection*{Velocity profile}
\begin{figure}
\includegraphics[width=\linewidth]{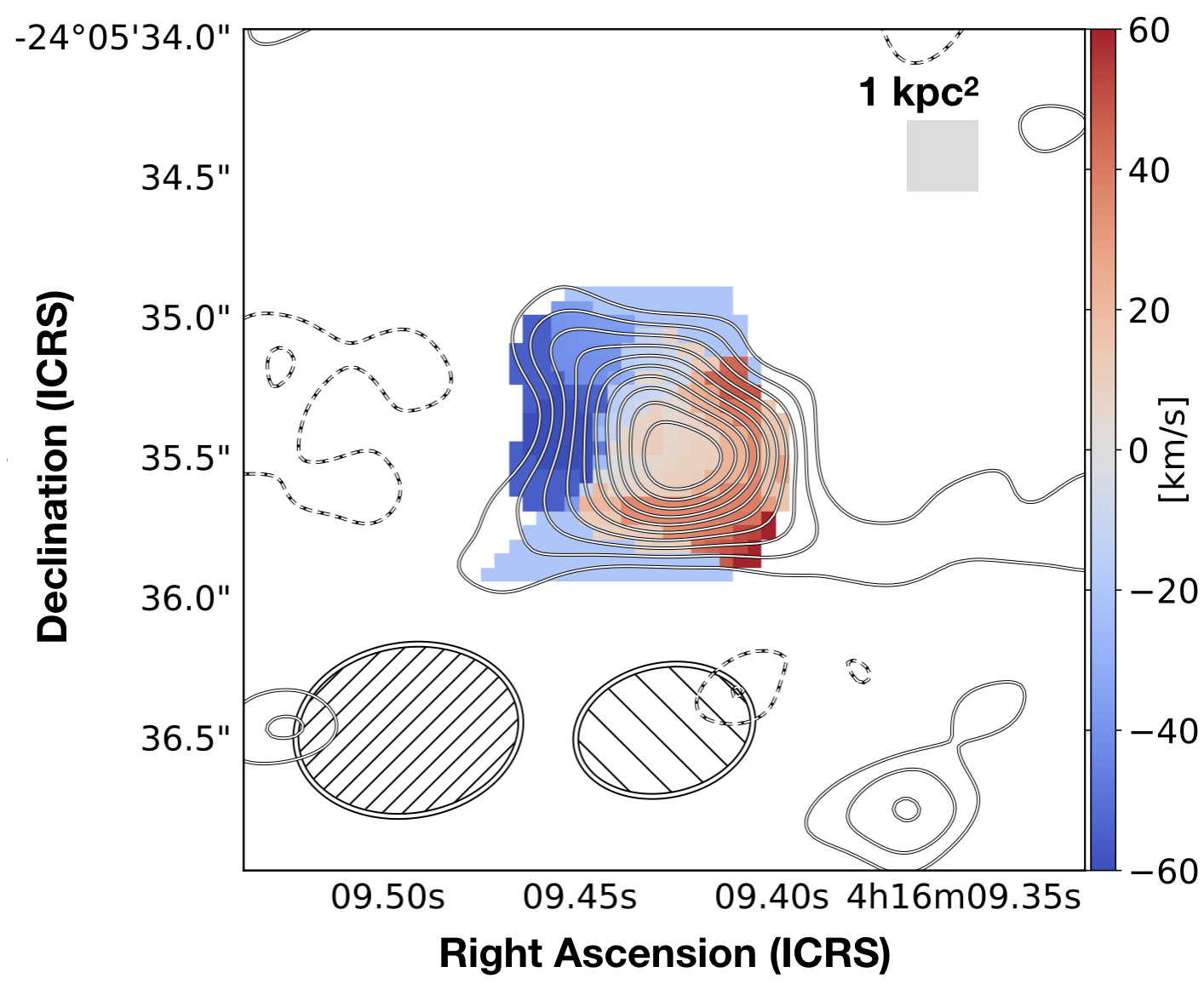}
\caption{The velocity component (first moment map) of the [C\,{\sc ii}] emission of MACS0416\_Y1 in the image plane shows a velocity gradient $\Delta$v of around 100 to 120 km/s. 
The foreground contours show the [C\,{\sc ii}] emission drawn at $-$2, 2, 3, ..., 11 and 12 $\sigma$, where the negative contours are indicated with dashed lines. The left-most ellipse in the bottom of the figure indicates the background beam size, and the more central ellipse indicates the beam size of the foreground contours. In the top-right, we indicate the size of one square kiloparsec in the source plane. If we interpret this velocity gradient as a rotation, this suggests that MACS0416\_Y1 is a rotation-dominated system, rotating around the central [C\,{\sc ii}] position. Other possibilities include galactic inflows or outflows, and a merging scenario.}
\label{fig:velocityfield}
\end{figure}
Observationally, one can classify star-forming galaxies into either rotation- or dispersion-dominated systems by their kinematic profile. Typically, rotation-dominated galaxies have $\Delta$v/2$\sigma_{\rm{tot}}$ > 0.4 \citep{Forster2009}. Here $\Delta$v is the peak-to-peak velocity across the source, and $\sigma_{\rm{tot}}$ is the line standard deviation. The high signal-to-noise ratio of our [C\,{\sc ii}] emission allows us to look for a velocity gradient in the [C\,{\sc ii}] emission. Figure~\ref{fig:velocityfield} shows the velocity component of MACS0416\_Y1, \textcolor{referee}{created with the \texttt{immoments} task within CASA.} The peak-to-peak velocity is around 100 to 120 km/s. This results in $\Delta$v/2$\sigma_{\rm{tot}}$ $\approx$ 0.6 to 0.7, suggesting a rotationally-dominated disc, similar to what was seen by \cite{Smit2018}. They reported on two LBGs at $z$ $\sim$ 6.8, with masses around 1.4 to 1.7 $\times$ 10$^9$~M$_{\odot}$.

From this, we can calculate the dynamical mass, $M_{\rm{dyn}}$, if we assume the virial theorem,
\begin{equation}
    M_{\rm{dyn}} = C \frac{r_{1/2} \sigma^2_{\rm{line}}}{G}.
\end{equation}
Here, $r_{1/2}$ is the half-light radius, $G$ is the gravitational constant, and the factor $C$ depends on various effects, such as line-of-sight projection, galaxy mass and the contributions from rotational and random motions. This value ranges from 2.25 for an average value of known galactic mass distribution models \citep{Binney2008} to 6.7 for dispersion-dominated systems \citep{Forster2009}. Given that MACS0416\_Y1 is rotationally-dominated, we take $C$ to be 3.4 from \cite{Erb2006}, which is derived for an average inclination angle of the disk, assuming a rotating system. Since our system is rotationally-dominated, we take the deconvolved major axis as the half-light radius, equalling 1.15 $\pm$ 0.35 kpc. 

We find a dynamical mass of (1.2 $\pm$ 0.4) $\times$ 10$^{10}$ M$_{\odot}$, although we note that the uncertainty in $C$ is not accounted for, and could vary by a factor of two. The UV-to-FIR modeling in \cite{Tamura2019} found a young stellar mass of $\sim$2.4$^{+0.5}_{-0.3}$ $\times$ 10$^{8}$ M$_{\odot}$ (varying slightly from model to model), and their analytic galaxy evolution model, where dust mass evolution is taken into account, predicts an older stellar component with a mass of M$_{\rm{star}}$ $\sim$3 $\times$ 10$^9$ M$_{\odot}$ \citep{Asano2013}. This places the stellar-to-dynamical ratio around 25\%, which is similar to \cite{Smit2018}, and in line with the star-forming galaxies at $z \approx 2-3$ \citep{Erb2006,Gnerucci2011}. The value is an order-of-magnitude larger than seen for a $z$ = 7.15 LBG \citep{Hashimoto2019}, although they only take the mass of the young stellar component into account. 
The galaxy evolution model of \cite{Tamura2019} further predicts a gas mass of $\sim 1 \times 10^{10}$~M$_{\odot}$, which is similar to the value we find with the kinematics, suggesting the enclosed mass is baryon-dominated.

Potentially, this apparent rotation could also be explained by other phenomena. For instance, the velocity gradient could also be a bi-polar outflow. This velocity gradient is in the same direction as the spur-like feature seen in the [C\,{\sc ii}] emission, although it is along the major axis of the UV-image of the LBG. The distance of the peak positive velocity relative to the centre of the [C\,{\sc ii}] emission is 3\,kpc, not accounting for projection and lensing magnification effects. There, the typical escape velocity for a halo with a mass of $\sim 1 \times 10^{10}$~M$_{\odot}$ is around 85\,km/s, which is of the same order as the peak positive velocity ($\sim$60\,km/s) enabling the possibility that the spur-like profile is a galactic outflow. This would agree with the feedback/outflow phase seen in simulations by \cite{Arata2019}. Gas column densities in the outflowing regions of the simulated galaxies by \cite{Arata2019} range between 1 and 100 M$_{\odot}$/pc$^2$. Our current beam-width does not allow us to resolve the individual outflowing regions, however an outflow moving at 60 km/s stretching over 1 square kpc (indicated in Figure~\ref{fig:velocityfield}) with a gas column density of 10 M$_{\odot}$ (typical in the range from \citealt{Arata2019}) would result an outflow-rate of 60 M$_{\odot}$/yr. The outflow rate could thus equal or surpass the star-formation rate (i.e. a mass-loading factor of around or greater than unity), impacting the galaxy evolution at high redshift.
There also exists observational indications for outflows at slightly lower redshifts, $z \sim 4 - 7$, which could support the outflow hypothesis for MACS0416\_Y1. [C\,{\sc ii}]-identified outflows have been suggested for $z = 5 - 7$ star-forming galaxies (SFR between 10 and 70 M$_{\odot}$/yr) in \cite{Fujimoto2019} by stacking the ALMA data of 18 galaxies. Similarly, among a sample of 118 normal star-forming galaxies at $4<z<6$, \cite{Ginolfi2019} reports high-velocity tails ($\sim$ 500~km/s) in the stacked [C\,{\sc ii}] emission, observed by ALMA. These velocity tails are indicative of outflows, and appear more prominent among more actively-starforming galaxies ($>25$\,M$_{\odot}$/yr). This correlation confirms the star formation-driven nature of the outflows, however these [C\,{\sc ii}]-line velocities are much larger than the $\sim$60~km/s seen in MACS0416\_Y1.

The reverse scenario, the inflow of gas, could similarly explain the observed phenomena. However, at $z$ = 8.31, the pristine inter-galactic medium is unlikely to be enriched enough to be bright in [C\,{\sc ii}] emission. Instead, it could be previously-ejected gas falling back in onto the galaxy, previously enriched by the older stellar population \citep{Tamura2019}. This would place the galaxy at the end of the feedback/outflow phase seen in the simulations by \cite{Arata2019}, and would suggest an imminent starburst phase. In this scenario, the free-fall timescale would be around 9~Myr, assuming a total enclosed mass of 1~$\times{}$~$10^{10}$~M$_{\odot}$, and a radius of 2.3~kpc. The high dust temperature (see Section \ref{sec:15mmnondetection}) does seem to contradict a late stage feedback/outflow scenario, since the warm dust ($>$80\,K) in the centre of the galaxy is expected to cool down efficiently, exacerbating the problem of the high dust temperature seen for MACS0416\_Y1. 

A final possibility is that this galaxy actually consists of multiple merging sources, smoothed over the beam size. This does agree with the multi-component nature seen in the rest-frame UV Hubble imaging. Currently, the most likely scenario seems a rotation-dominated disk given the observed velocity gradient, although future high-resolution observations of MACS0416\_Y1, or a stacked spectrum of multiple $z>8$ LBGs, are required to confirm any of the above scenarios.

\subsubsection*{[C\texorpdfstring{\,{\sc ii}}{II}] geometry}
\begin{figure*}
\includegraphics[width=\linewidth]{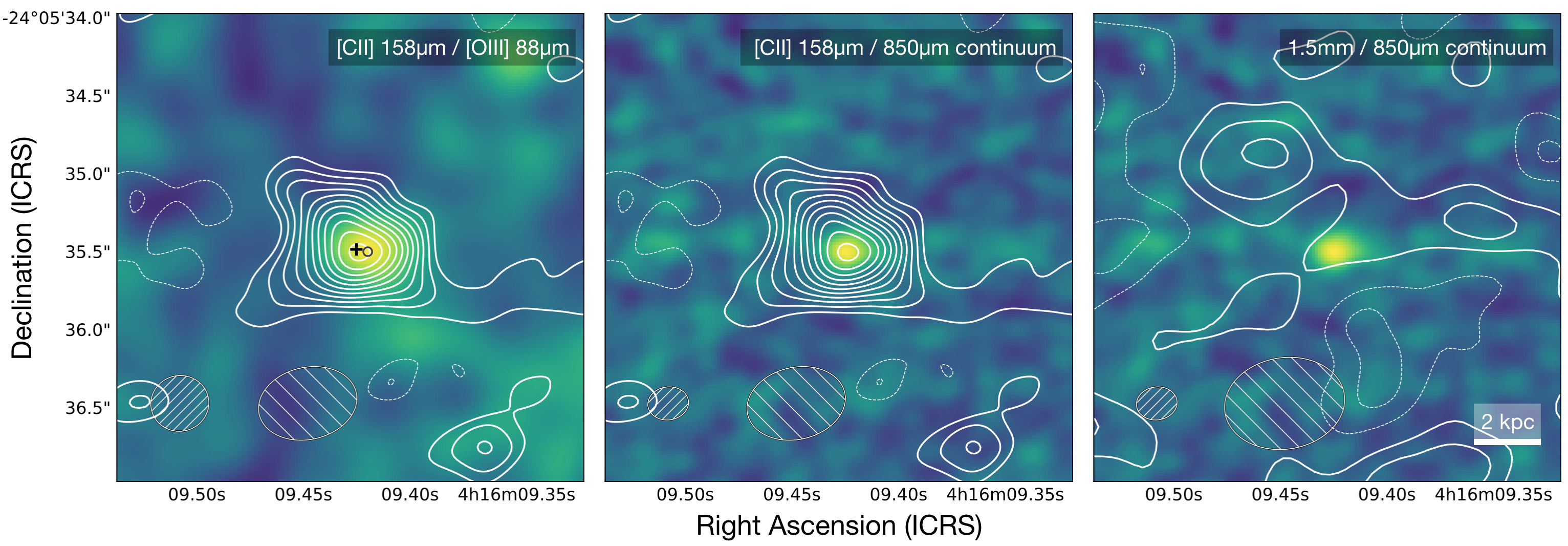}
\caption{\textit{Left and Middle}: The [C\,{\sc ii}] contours (drawn at $-$2, 2, 3 ... 11 $\sigma$, where negative contours are indicated with dashed lines) are shown on top of the [O\,{\sc iii}] and 850 $\mu$m dust continuum emission images from \citet{Tamura2019}, respectively. There appears to be a positional offset between the [C\,{\sc ii}] and [O\,{\sc iii}] position, agreeing with the eastern and center component seen in the Hubble images, and indicated with a plus ([C\,{\sc ii}]) and circle ([O\,{\sc iii}]). The [C\,{\sc ii}] and 850~$\mu$m dust continuum emission profiles appear to be aligned, suggesting the origin of the radiation to be in the same location. \textit{Right}: The 1.5 mm dust continuum contours are drawn at more levels for extra contrast, namely $-$2,$-$1, 1, 2, 3 $\sigma$, with negative contours indicated with dashed lines, and are shown on top of the continuum emission image at 850 $\mu$m from \citet{Tamura2019}. No significant emission can be seen, which places a strong upper limit on the LBGs spectrum. We discuss the relevance of this in Section \ref{sec:15mmnondetection}. The left-most ellipse indicates the background beam size, and the more centrally-located ellipse indicates the beam size of the contours.}
\label{fig:OIIIandDUSTcomparison}
\end{figure*}
The left panel of Figure \ref{fig:OIIIandDUSTcomparison} shows the [C\,{\sc ii}] emission as contours on top of the previously-observed [O\,{\sc iii}] emission image. There appears to be a slight angular offset between the [C\,{\sc ii}] and [O\,{\sc iii}] emission, which could be due to the different place of origin. [C\,{\sc ii}] predominantly emerges from PDRs, whereas [O\,{\sc iii}] only exists within H\,{\sc ii} regions. 

The exact position of [O\,{\sc iii}] from \cite{Tamura2019} is ($\alpha_{\rm{ICRS}}$, $\delta_{\rm{ICRS}}$) = (04$^{\rm{h}}$16$^{\rm{m}}$09\fs420, $-$24$^{\circ}$05$^{{\rm{'}}}$35\farcs495), which is 95 milli-arcseconds offset from the [C\,{\sc ii}] emission, \textcolor{referee}{where both positions are measured by fitting a 2D Gaussian shape to the line emission.} The systematic positional uncertainty in the astrometry for the [C\,{\sc ii}] and [O\,{\sc iii}] emission are around 20 and 6 milli-arcseconds, respectively. Combined with the positional uncertainties, this is a 2$\sigma$ offset, and if real, would equal a projected distance of $\sim$450 parsec. 

Similar angular offsets have been seen both in previous observations and in simulations. \cite{Carniani2017} find more extended [C\,{\sc ii}] emission, and a $\sim$1\,kpc offset between the brightest [C\,{\sc ii}] and [O\,{\sc iii}] clump, for a normal star-forming galaxy at $z$ = 7.1. They find both the [C\,{\sc ii}] and [O\,{\sc iii}] emission are significantly offset ($\sim$3 to 5\,kpc) from the main rest-frame UV emission, find three [O\,{\sc iii}] emission clumps with distinct relative velocities, and find the [C\,{\sc ii}] emission consisting of a clumpy and an extended component.
Recent simulations by \cite{Katz2019} show the positional distribution of infrared lines over a galaxy, and predict an offset between [C\,{\sc ii}] and [O\,{\sc iii}] due to their different clouds of origin \citep{Carilli2013} at very high redshift (z $\sim$ 10). In one execution of their simulation, they find angular offsets between the peaks of [C\,{\sc ii}] and [O\,{\sc iii}] emission of 0.5 to 1 kpc. Similarly, recent work by \cite{Pallottini2019} sees a positional ($\sim$1~kpc) and spectral ($\sim$100~km/s) offset between [C\,{\sc ii}] and [O\,{\sc iii}] emission in their simulated galaxies. They show that this is both because [C\,{\sc ii}] and [O\,{\sc iii}] originate from different environments in the ISM, and because a more extended, but inhomogenous, distribution of [C\,{\sc ii}] emission can shift the peak flux away from the peak [O\,{\sc iii}] emission. 

While we can expect a difference in velocity and velocity width of the [C\,{\sc ii}] and [O\,{\sc iii}] lines, the errorbars in Figure \ref{fig:Linefit35.pdf} show that the spectral lines do not appear to differ significantly. Quantitatively, the peak of each line is offset by 15 $\pm$ 15 km/s, and the velocity width between the lines differs by 50 $\pm$ 36 km/s. 

The middle image of Figure \ref{fig:OIIIandDUSTcomparison} suggests the [C\,{\sc ii}] emission originates in the same region as 850~$\mu$m continuum emission. This is similar to what was seen in \cite{Pallottini2019}, and in agreement with the statistical relations seen both for high-redshift sources (e.g. \citealt{Smit2018}) and sub-mm sources in general (e.g. \citealt{deLooze2014}).

\subsection{The lack of 1.5~mm continuum emission}
We had expected to detect the 160~$\mu$m rest-frame emission from MACS0416\_Y1, given the assumed spectrum from \cite{Tamura2019}. Here, we analyse the significance of the non-detection, and provide explanations for the lack of continuum emission. 
\label{sec:15mmnondetection}
\label{sec:dis_cont}
\subsubsection*{Fitting a spectrum}
\begin{figure}
\includegraphics[width=\linewidth]{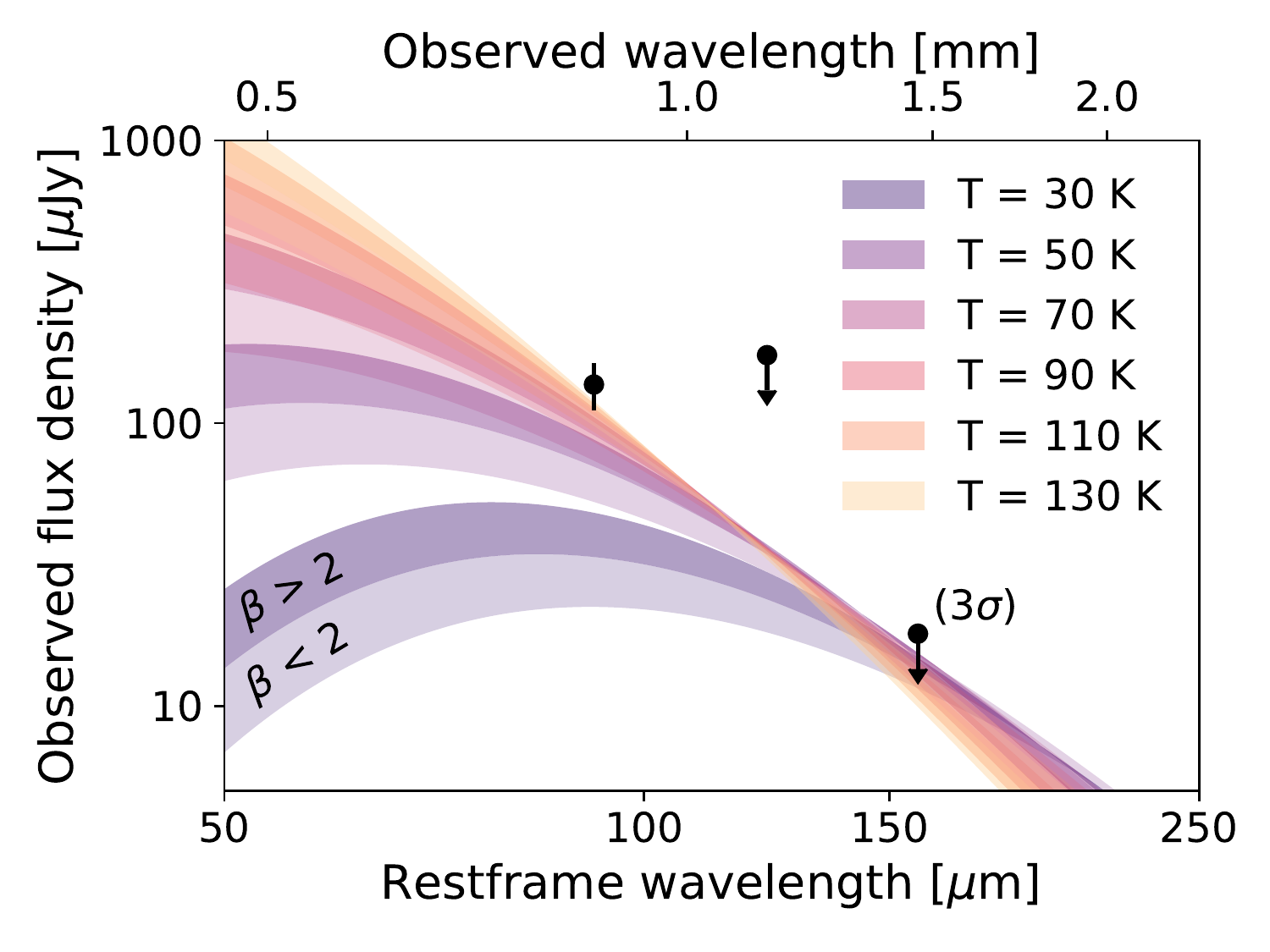}
\caption{We fit modified blackbodies with various fixed temperatures to the single continuum data point, and two upper limits (shown at 3$\sigma$), with $\beta$ varying from 1.5 to 2.5. The lighter fills show the modified blackbody spectra with 1.5 $<$ $\beta$ $<$ 2.0, while the darker fills show the spectra with 2.0 $<$ $\beta$ $<$ 2.5. Our non-detection of continuum at 1.5~mm (rest-frame $\sim$160~$\mu$m) provides a stringent lower-limit on the dust temperature. We account for both CMB-heating and CMB-contrast using the equations of \citet{daCunha2013}, in order to discuss the CMB-corrected dust temperature (T$_{z = 0}$) in the observed flux density.}
\label{fig:SED}
\end{figure}
\begin{table*}
\caption{The fitting parameters of the tested single-temperature spectrum fits}
\label{tab:Tfits}
\centering
\resizebox{\linewidth}{!}{\begin{tabular}{llllllllllll}  
\hline \hline
& \multicolumn{3}{c}{$\beta$ = 1.5} & & \multicolumn{3}{c}{$\beta$ = 2.0} & & \multicolumn{3}{c}{$\beta$ = 2.5} \\ \cline{2-4} \cline{6-8} \cline{10-12}
T$_{z = 0}$ & $\mu$IR Lum. &  $\chi^2$ & $\mu$M$_{\rm{dust}}$ & T$_{z = 0}$ & $\mu$IR Lum. & $\chi^2$ & $\mu$M$_{\rm{dust}}$ & T$_{z = 0}$ & $\mu$IR Lum. &  $\chi^2$ & $\mu$M$_{\rm{dust}}$	 \\ 
(K) & (10$^{11}$ L$_{\odot}$) & - & (10$^{6}$ M$_{\odot}$) & (K) & (10$^{11}$ L$_{\odot}$)  & - & (10$^{6}$ M$_{\odot}$) & (K) & (10$^{11}$ L$_{\odot}$) & - & (10$^{6}$ M$_{\odot}$)\\
\hline
30 & 0.31 & 18.9  	& 11	& - & 0.49 		& 15.8  	& 5.1 & - & 0.76 & 12.2 & 2.28	    \\
50 & 1.34 & 10.3  	& 3.0	& - & 2.21 		& 6.79  	& 1.2 & - & 3.56 & 3.88 & 0.47	    \\
70 & 4.53 & 6.15  	& 1.6	& - & 8.11 		& 3.40 	 	& 0.6 & - & 14.4 & 1.51 & 0.22	    \\
90 & 12.2 & 4.15  	& 1.1	& - & 23.8 		& 2.00 	 	& 0.4 & - & 46.4 & 0.64 & 0.14	    \\
110 & 28.5 & 3.07  	& 0.8	& - & 59.8 		& 1.30   	& 0.3 & - & 126.8 & 0.23 & 0.10	\\
130 & 59.2 & 2.42  	& 0.7	& - & 133.4 	& 0.90   	& 0.2 & - & 310.8 & 0.019 & 0.085	\\
\hline
121 & 46.0 & 2.7 \textbf{(90\%)} & 0.74 & 80 & 15.9 & 2.7 \textbf{(90\%)} & 0.5 & 60 & 8.95 & 2.7 \textbf{(90\%)} & 0.35 \\ \hline
\end{tabular}}
\raggedright \justify \vspace{-0.2cm}
\textbf{Notes:} For each value of $\beta$, reading from the left, the columns are: Column 1 - CMB-corrected dust-temperature; Column 2 - Far-infrared luminosity, not corrected for magnification, $\mu$; Column 3 - $\chi^2$ value, measuring quality of fit; Column 4 - Dust mass, not corrected for magnification. The bottom row shows the 90\% lower limit on the dust temperature, far-infrared luminosity and dust mass for each $\beta$, found by interpolating the $\chi^2$ to a value of 2.7 \citep{Avni1976}.
\end{table*}
The basic theoretical description of a sub-mm spectrum of a dusty source is determined exclusively by the dust temperature and the dust-emissivity index, $\beta$ (e.g. \citealt{Dunne2001}). At higher redshifts, the CMB becomes an important component in the heating of the dust. Throughout this analysis, we use the adjustment equations from \cite{daCunha2013} to include the effects of CMB, both accounting for the CMB dust heating and the contrast of the emission against the CMB. This allows us to discuss CMB-corrected dust temperatures and observed flux densities, as though the source were affected by the CMB at redshift 0 (i.e. T$_{z = 0}$, with T$_{\rm CMB}$ = 2.73~K).

In Figure \ref{fig:SED}, we show the results from fitting a single-temperature modified blackbody with different $\beta$ for a range of possible temperatures. We fit these modified blackbodies by adjusting the normalization, $A$, such that we minimize the chi-squared, $\chi^2$. Our data are the single 850 $\mu$m (rest-frame 90 $\mu$m) continuum detection and two continuum upper limits at 1.1 and 1.5 mm (rest-frame 120 and 160~$\mu$m). The 120 $\mu$m non-detection is from a Band 6 ALMA project (P.I. F. Bauer, 2013.1.00999.S) aiming to constrain the photometry of sources behind strong-lensing clusters \citep{GL2017}. The 160~$\mu$m non-detection, with an r.m.s. noise down to 6 $\mu$Jy, provides a significant upper-bound on the dust parameter space probed by these observations.
The upper limits are taken into account as discussed in \cite{Sawicki2012}, where an adjustment on the $\chi^2$ is demonstrated to be
\begin{equation}
\delta \chi^2 = -2 \ln \int_{-\infty}^{\sigma_j + S_j} \exp\left[- \frac{1}{2} \left( \frac{f - A S_{model}}{\sigma_{j}} \right)^2 \right] df.
\end{equation}
In this equation, we calculate $\chi^2$ for over all non-detections j, and integrate the Gaussian flux distribution along flux, $f$, up to the detection criterion, which is set to the measured standard-deviation ($\sigma_j$) plus the value observed at the central position ($S_j$). The adjustment to the $\chi^2$ measures the probability that the model-predicted value ($A S_{model}$) is scattered below the detection limit due to random fluctuations.

The fitted spectra in Figure \ref{fig:SED} appear to favour higher temperatures and $\beta$-values, and the 1.5~mm continuum upper-limit seems to exclude any temperature less than $\sim$90~K for $\beta$ < 2, although for increasing $\beta$ > 2, slightly cooler temperatures are possible.

Table \ref{tab:Tfits} quantifies the results, by showing the $\chi^2$ value for different fixed temperature and $\beta$ values, which we use as a measure of the fitting quality. Given the degeneracy between $\beta$ and temperature, we cannot fit these values directly, as this fit results in physically-unrealistic temperatures (e.g. T$_{\rm dust}$ $>$ 200\,K or $\beta$ $>$ 4). Instead, we find the 90\% confidence limit from the $\chi^2$ value. From \cite{Avni1976}, we use that a 90\% confidence limit can be placed on a single varied parameter if the fitting of a model results in a $\Delta \chi^2$ value of 2.7. For each value of $\beta$, we calculate the 90\% lower limit for the dust temperature, which we show in the final row of the table. We linearly-interpolate the temperature for the two nearest $\chi^2$ values to 2.7 for each value of $\beta$. This calculates the 90\% lower limit on the temperature, luminosity and dust mass. The resulting $\chi^2$ from the two constraining data points (at 850 and 1500 $\mu$m) agree with the visual analysis of Figure \ref{fig:SED}, suggesting dust temperatures $>$ 80~K for $\beta$ = 2. We estimate the dust mass according to equation 1 in \cite{Magdis2011},
\begin{equation}
M_{\rm dust} = \frac{S_{\nu}D_{\rm L}^2}{(1+z)\kappa_{\rm rest}B_{\nu}(\lambda_{\rm rest},T_d)}.
\end{equation}
Here, S$_{\nu}$ is the observed flux density, $D_{\rm L}$ is the luminosity distance, $z$ is the spectroscopic redshift, $\kappa_{\rm rest}$ is the rest-frame dust mass absorption coefficient at the observed wavelength and $B_{\nu}(\lambda_{\rm rest},T_d)$ is the black-body radiation expected for a $T_d$ [K] source at $\lambda_{\rm rest}$. The dust emissivity is assumed such that $\kappa_{\rm rest}$ = $\kappa_{\rm 850 \mu m}$($\nu$ /$\nu_0$)$^{\beta}$, where $\nu_0$\,=\,353 GHz, and $\kappa_{\rm 850 \mu m}$ = 0.15\,m$^2$/kg \citep{Draine2003}.
The black-body emission is corrected for its contrast against the CMB, and the dust temperature is corrected for the heating by the CMB.

We note that the dust temperature has a significant impact on the dust mass. The magnified dust mass estimate in \cite{Tamura2019} from Band 7 ALMA observations is 5.6 $\times$ 10$^6$~M$_{\odot}$. This dust mass is very temperature-dependent, where even the lower-limit on the dust temperature suggests magnified dust-masses of around 0.5~$\times$~10$^6$~M$_{\odot}$, a ten-fold decrease in dust mass from the estimate in \cite{Tamura2019}. The effect of the dust emissivity coefficient, $\beta$, on the dust mass is less strong, changing the dust mass by $\sim$50 per cent. We discuss the consequences of the variance in dust mass with dust temperature below.

\subsubsection*{Comparison of [C\texorpdfstring{\,{\sc ii}}{II}] emission to dust temperature}
\begin{figure*}
\includegraphics[width=\linewidth]{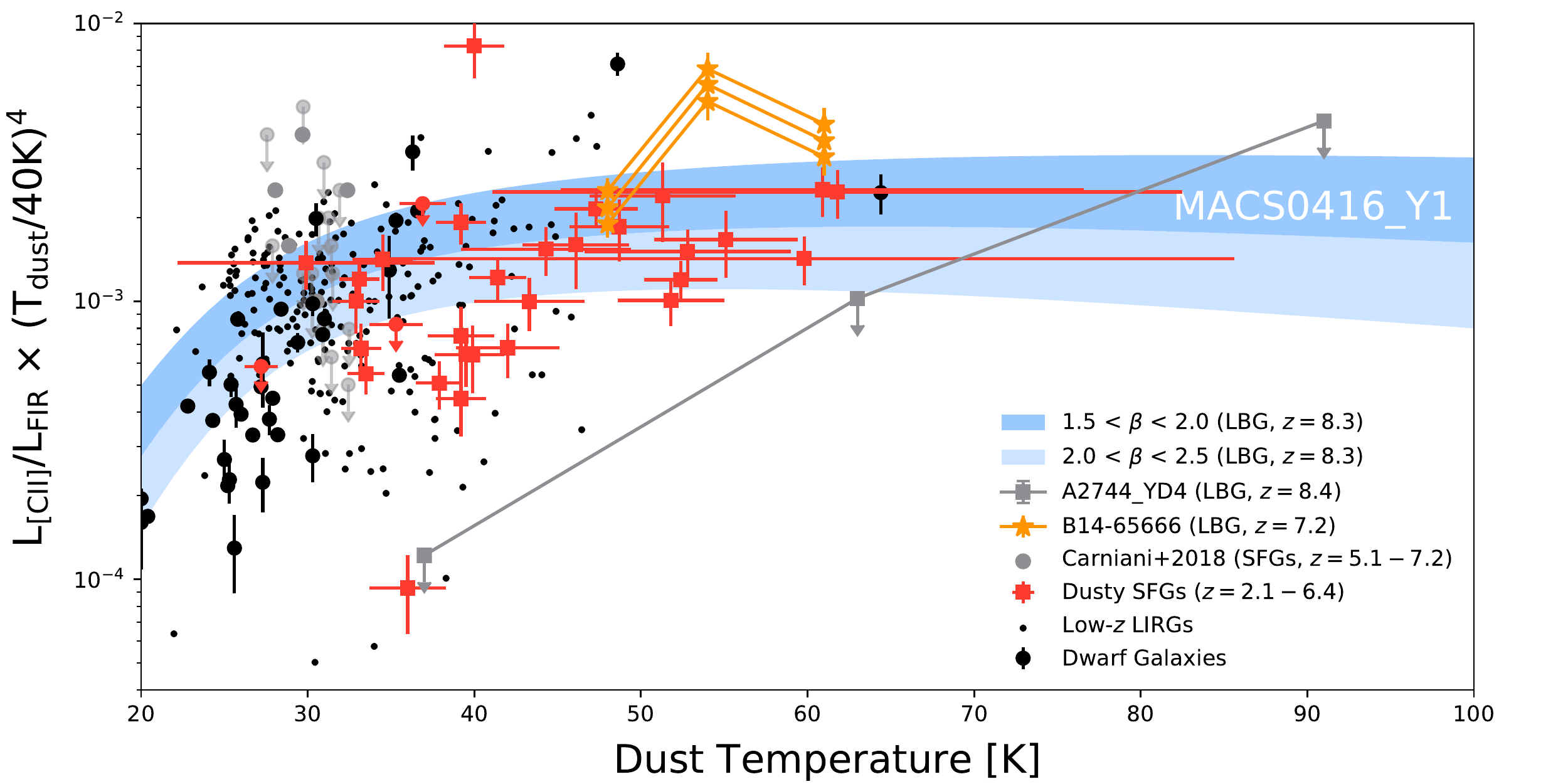}
\caption{The temperature-corrected [C\,{\sc ii}] to far-infrared luminosity ratio is shown as a function of the dust temperature. We compare a range of temperatures and $\beta$s of MACS0416\_Y1 against high-redshift LBGs, high-redshift dusty star-forming galaxies, low redshift galaxies and nearby dwarf galaxies. MACS0416\_Y1 appears to have a similar ratio as the other sources, regardless of the dust temperature, and the ratio is only mildly dependent on $\beta$.}
\label{fig:Gullberg.pdf}
\end{figure*}
We show a modified version of the [C\,{\sc ii}] to far-infrared luminosity ratio as a function of temperature for MACS0416\_Y1 in Figure~\ref{fig:Gullberg.pdf}. Since the luminosity is very temperature-dependent, we use the temperature-corrected luminosity in order to compare the [C\,{\sc ii}] to far-infrared luminosity ratio for MACS0416\_Y1 over a large range of temperatures (20 - 100~K, \citealt{Malhotra1997,Gullberg2015}). We calculate this temperature-corrected ratio by multiplying the [C\,{\sc ii}] to far-infrared luminosity ratio by the temperature normalised at 40\,K to the fourth power.\footnote{\textcolor{referee}{We choose 40\,K, similar to \cite{Gullberg2015}, since this is a typical dust temperature in the local Universe, which ensures that the temperature-corrected luminosity ratios are of the same order as non-corrected ratios.}}

We compare our single-temperature fit of MACS0416\_Y1 to high- and low-redshift sources. The temperatures of all the high-redshift galaxies are derived from either one or two wavelength observations. \cite{Hashimoto2019} provide the luminosity for three different $\beta$ and temperature values ([T$_{\rm dust}$, $\beta$] = [48~K, 2.0], [54~K, 1.75], and [61~K, 1.5]) for the entire source B14-65666 ($z = 7.15$ LBG), and for the two separate components. 
The upper-limit on [C\,{\sc ii}] emission of A2744\_YD4 ($z = 8.38$ LBG) is shown for three different temperatures, since modeling by \cite{Behrens2018} suggests a significantly higher dust temperature (93~K) than the initial observations of \cite{Laporte2017} indicate (37 - 63~K). Furthermore, an ALMA Band-5 upper limit \citep{Laporte2019} suggests a 3$\sigma$ dust temperature lower-limit of 43~K.
\cite{Carniani2018} report the dust temperature, far-infrared luminosity, and [C\,{\sc ii}] emission of 5 sources (and 16 upper limits) at $z$ = 5 to 7. The luminosity was calculated for a 30~K, $\beta$ = 1.5 single-temperature modified black-body, although for graphing purposes we plot them with temperatures ranging from 25~K to 35~K. Whilst the temperature in their analysis is fixed, the temperature-corrected luminosity ratio is similar across the temperature range, given the flatness of the \textit{blue fill}, especially for higher temperatures.
We show several dusty star-forming galaxies (DSFGs) at 2.1 $<$ $z$ $<$ 6.4. These consist of twenty sources detected with the South Pole Telescope (SPT), and fourteen other DSFGs selected from various sub-mm/mm surveys \citep{Gullberg2015}.
The low-redshift luminous infrared galaxies (LIRGs) are from the Great Observatories All-Sky LIRG Survey (GOALS) sample \citep{Armus2009,DiazSantos2013}, where we calculate the temperature using Eq.\,2 in \cite{DiazSantos2017}. 
The dwarf galaxies are from the Herschel Dwarf Galaxy Survey \citep{Madden2013,Cormier2015}.

We find a good agreement between the temperature-corrected [C\,{\sc ii}]-to-FIR luminosity ratio for MACS0416\_Y1 compared to other sources over a large range of potential dust temperatures. Our agreement with the previously-observed data does not depend on $\beta$, as adjusting the $\beta$ from 1.5 to 2.5 only changes the luminosity ratio by a factor of $\sim$3. 
The graph further shows that the non-detection of A2744\_YD4 \citep{Laporte2019} does not appear to require a significantly-lower [C\,{\sc ii}] luminosity than expected, especially when the temperature would be greater than 60~K, as a simulation in \cite{Behrens2018} indicated.

\subsubsection*{High $\beta$ or high dust temperature?} \label{sec:highTorhighBeta}
Table \ref{tab:Tfits} suggests either a high dust emissivity index ($\beta>2$) or dust temperature ($T>80$~K). Typical ULIRGs appear to have $\beta<2$ \citep{Clements2018}, and the Milky Way has an average $\beta$ around 1.6 \citep{Planck2013}. Similarly, only a few of the galaxies in Figure \ref{fig:Gullberg.pdf} have dust temperatures beyond 50\,K, and temperatures beyond 70\,K appear out of the ordinary for anything but the recent simulation by \cite{Behrens2018} (e.g. \citealt{Faisst2017}).

The high-redshift nature of MACS0416\_Y1 will almost certainly result in a different chemical composition of the ISM due to the restricted number of elemental pathways for nucleosynthesis, when compared to low-redshift galaxies that allow for pathways that require more time, such as lower-mass stars (e.g. \citealt{Maiolino2019}). A different composition of dust grains could have caused a high $\beta$ in MACS0416\_Y1. For example, \cite{Demyk2013} finds dust grain $\beta$ values ranging from 0.8 to 3, depending strongly on the specific chemical composition of the dust grains. They provide an overview of the physics of dust emission, studied both with theoretical models and in laboratory experiments. Furthermore, the dust has to, almost exclusively, rely on supernovae for its origin. At $z=8.3$, the first stars have only been around for 500~Myr \citep{Tegmark1997}, so the time for 0.6\,-\,8 M$_{\odot}$ stars to reach their AGB phase is restricted ($>$100 Myr), and dust accretion growth is hindered by photo-ionization, the low metallicity and the restricted time. 

MACS0416\_Y1 would not be the first observed galaxy with a $\beta$ $>$ 2. For example, ALMA observations by \cite{Kato2018} indicate a $\beta$ = 2.3~$\pm$~0.2 in a Ly-$\alpha$ blob at $z$ = 3.1. They suggest their high $\beta$ is due to the chemical composition \citep{Demyk2013}. In the local Universe, \cite{Smith2012} finds a high $\beta$ ($\sim$3) in the galaxy core of Andromeda, and discusses its origin due to grain coagulation or icy mantles on the surface of grains in denser regions.

The dust produced in the initial episode of star formation in MACS0416\_Y1, at around $z$ $\sim$ 15 \citep{Tamura2019}, could have been formed by Population\,{\sc iii} stars \citep{Nozawa2014}.
\cite{deRossi2018} suggest that this interstellar dust created by Population\,{\sc III} stars would be silicate-rich, and the resulting far-infrared spectrum would appear substantially warmer than expected. They are able to match their predicted SED to Haro 11, a local, moderately low-metallicity galaxy ($\sim$0.25~Z$_{\odot}$) undergoing a very young and vigorous starburst. These conditions suggest this galaxy approximates the relevant conditions for young Population\,{\sc III} stars at high redshift. The existence of Population\,{\sc III} stars has not been directly confirmed, however, and still remain controversial (e.g. \citealt{Sobral2015,Sobral2019,Bowler2017,Shibuya2018}).

Of course, our observations of MACS0416\_Y1 can only probe the dust that emits light, and therefore they probe the luminosity-weighted dust temperature \citep{Scoville2016}. Modeling by \cite{Behrens2018} showed that this might significantly differ from the mass-weighted dust temperature distribution, especially for high redshift sources. For the $z$ = 8.38 source A2744\_YD4, they find the hot dust to be mostly situated around compact young stellar clusters, whose strong interstellar radiation fields heat the dust efficiently due to the small optical depth. In their case, only 25\% of the dust is at T$_{\rm{dust}}$ > 60~K, but it contributes to more than 80\% of the luminosity. Thus, perhaps we are only witnessing a small portion of the dust that is exceptionally warm, with the majority of the dust existing at lower temperatures.

The size of the dust grains could further impact the dust temperature we see for MACS0416\_Y1. Since dust growth is potentially suppressed by photo-ionization and does not have the time to coagulate in the early Universe, dust grain sizes of high-redshift galaxies might be smaller than in the local Universe. \cite{Ferrara2017} find that the grain size could change the temperature of the dust, with an increase in average dust temperature from 45K for grains $>$0.1$\mu$m to 60~K for grains smaller than 0.1$\mu$m.

Computer models are also suggesting that the high dust temperatures in MACS0416\_Y1 are not an anomaly.
Simulations by \cite{Arata2019} also find that dust temperatures of $z \sim 6$ galaxies could increase to $\sim$70~K, and that the star-forming central region of their simulated galaxy appears to have the hottest dust. Their analysis shows a time-variable far-infrared luminosity, and with it the temperature. Due to stellar feedback processes, the simulated galaxy oscillates between a dust-obscured star-forming galaxy and UV-dominated galaxy with a reduced star-formation rate. They also report on the difference between the mass- and luminosity-weighted dust temperature, and find a similar result as \cite{Behrens2018}. The analytical model of primordial dust emission from \cite{deRossi2019} also finds the central region to become brighter, relative to the total luminosity, for higher redshift galaxies. High redshift systems are more concentrated (i.e. higher dust densities in the centre), enhancing the heating efficiency by stellar radiation, and increasing the luminosity-weighted temperature.

Finally, the high temperature in MACS0416\_Y1 agrees with the high [O\,{\sc iii}]-to-[C\,{\sc ii}] ratio we find, as well as the relatively-low metallicity ($\sim$0.2~Z$_{\odot}$) inside this galaxy. The correlation between dust temperature and [O\,{\sc iii}]-to-[C\,{\sc ii}] ratio was also seen in observations by \cite{DiazSantos2017} and \cite{Walter2018}, who show a correlation between this ratio and the dust temperature at both low and high redshift. \cite{Faisst2017} finds an anti-correlation between the metallicity and dust temperature, by studying a selection of lower-redshift analogues of $z>5$ galaxies with low metallicity. This might not be a fair comparison, since (i) the dust temperatures of these sources do not exceed 60\,K, (ii) the most significant high-redshift source in \cite{Walter2018} is a quasar, which could potentially influence the heating mechanisms and the origin of the [O\,{\sc iii}] emission, and (iii) the [O\,{\sc iii}]-to-[C\,{\sc ii}] ratios of all sources in these studies range from 0.05 to 2. 

\subsubsection*{Astronomical consequences}
Our investigations of MACS0416\_Y1 have only targeted the 90 and 160~$\mu$m rest-frame emission, as the [O\,{\sc iii}]~88$\mu$m and [C\,{\sc ii}]~158$\mu$m spectral lines are the only available bright lines visible in the atmospheric windows. 
Due to the long integration times required to detect MACS0416\_Y1 and other high-redshift sources, most studies present a far-infrared luminosity with a single assumed temperature (e.g. \citealt{Watson2015} and \citealt{Carniani2017}). 
The luminosity varies significantly as a function of temperature, which not only poses a problem for the analysis in the previous section, but also for astronomical studies in general.

One of the main conclusions from \cite{Tamura2019} is the existence of an older stellar population in MACS0416\_Y1, determined using a dust evolution model to explain its dust mass and metallicity. The increase in dust temperature, combined with the potential change in $\beta$, can increase the per-mass luminosity of dust by 100-fold, as can be seen in Table \ref{tab:Tfits}, allowing the dust temperature to significantly impact the study of high-redshift objects in general.
Moreover, the amount of dust seen in MACS0416\_Y1 and other high-redshift galaxies (e.g. \citealt{Hashimoto2019, Laporte2019}) suggests an extreme dust-production scenario, where SNe have to produce dust at their maximum observed efficiency to reproduce the previously-observed dust masses \citep{Lesniewska2019}. 
The increase in dust temperatures reduces the required dust masses, and significantly relaxes the constraints on the dust production mechanisms at high redshift. 

\textcolor{referee}{For MACS0416\_Y1 in particular, the decrease in dust mass is not expected to significantly influence the age of the older stellar component ($\sim$300~Myr). 
While both the high metallicity and high dust mass of MACS0416\_Y1 seemed to require a high age \citep{Tamura2019}, lowering the dust mass would still leave an intact metallicity-based lower limit on the age. However, determining the effects of a decrease in dust temperature on the older stellar population requires extensive modelling using stellar population synthesis models and dust production mechanisms, which is beyond the scope of this paper.}

The reduction of dust mass in MACS0416\_Y1 and other high-redshift galaxies due to increasing dust temperatures and/or $\beta$, however, does not mean that the dust-obscured star-formation becomes less important. Arguably, it becomes more important, as the dust spectrum would peak at shorter wavelengths ($\lambda_{\rm{peak}}$ $\approx$ 40 - 70 $\mu$m), which still remain largely unprobed in the high-redshift Universe, both due to the lack of sensitive instrumentation (as well as atmospheric opacity) in this wavelength range, and the lack of bright spectral lines in this wavelength range (although [O\,{\sc iii}] at 52$\mu$m and [O\,{\sc i}] at 63$\mu$m are available). These spectral lines are often the main observational goal, where the dust continuum detection occurs almost serendipitously. We therefore advise towards a more thorough examination of dust temperatures in the early Universe, as well as the need for future instrumentation that can probe the peak of warm dust in the Epoch of Reionization, such as the planned Origins Space Telescope\footnote{https://asd.gsfc.nasa.gov/firs/} mission \citep{Meixner2019}.

\section{Conclusions}
We have detected the [C\,{\sc ii}]-emission of an LBG at $z$ = 8.3116, and provide a stringent upper-limit on the 1.5 mm continuum emission. The [C\,{\sc ii}]-emission compared to [O\,{\sc iii}]-emission indicates strong interstellar radiation fields and/or a low PDR covering fraction, which is also seen for nearby dwarf galaxies with high [O\,{\sc iii}]-to-[C\,{\sc ii}] ratios. A spatial offset between the [C\,{\sc ii}]- and [O\,{\sc iii}]-emission suggests that the emission originates from a different part of the ISM. The [C\,{\sc ii}]-emission traces the previously-detected 850 $\mu$m dust continuum, suggesting it originates from the same ISM. The far-infrared to [C\,{\sc ii}] luminosity ratio is in line with previously found ratios for sub-mm galaxies, suggesting a normal dusty ISM, when corrected for dust temperature. 
The velocity gradient across MACS0416\_Y1 suggests it is a rotationally-dominated, with a low stellar-to-dynamical gas mass ratio, although we also discuss the potential of an outflow or inflow of gas. 

No dust-continuum was detected at 1.5~mm down to 18 $\mu$Jy (3$\sigma$). Combined with the 137 $\mu$Jy dust continuum at 850 $\mu$m, this indicates a significantly warmer dust component than local or mid-redshift galaxies, and would also necessitate a very high value of dust emissivity index ($\beta$) of 2, or potentially higher. We further discuss the risks of basing a far-infrared luminosity on one or two continuum measurements, and advise towards probing the short-wavelength part of the dust continuum.

\section*{Acknowledgements}
This paper makes use of the following ALMA data: ADS/JAO.ALMA 2017.1.00225.S, 2016.1.00117 and 2013.1.00999.S. ALMA is a partnership of ESO (representing its member states), NSF (USA) and NINS (Japan), together with NRC (Canada), MOST and ASIAA (Taiwan), and KASI (Republic of Korea), in cooperation with the Republic of Chile. The Joint ALMA Observatory is operated by ESO, AUI/NRAO and NAOJ. \textcolor{referee}{We would like to thank Nicholas Laporte for his insightful comments.}
TB and YT acknowledge funding from NAOJ ALMA Scientific Research Grant Numbers 2018-09B and JSPS KAKENHI No. 17H06130. 
EZ acknowledges funding from the Swedish National Space Board.
TO is financially supported by MEXT 
KAKENHI Grant (18H04333) and the Grant-in-Aid (19H01931). T.H. was supported by Leading Initiative for Excellent Young Researchers, MEXT, Japan.

%%%%%%%%%%%%%%%%%%%%%%%%%%%%%%%%%%%%%%%%%%%%%%%%%%



%\bibliographystyle{mnras}

\bibliographystyle{mnras}
\bibliography{bibMake}

%%%%%%%%%%%%%%%%%%%%%%%%%%%%%%%%%%%%%%%%%%%%%%%%%%

%%%%%%%%%%%%%%%%% APPENDICES %%%%%%%%%%%%%%%%%%%%%

\appendix

%%%%%%%%%%%%%%%%%%%%%%%%%%%%%%%%%%%%%%%%%%%%%%%%%%

% Don't change these lines
\bsp  % typesetting comment
\label{lastpage}
\end{document}